\documentclass[11pt,a4paper,english,superscriptaddress,aps,preprint]{revtex4}
\usepackage{amsmath}
\usepackage{amssymb}
\usepackage{graphicx}
\makeatletter
\usepackage{babel}
\newcommand{\bea}{\begin{eqnarray}}
\newcommand{\eea}{\end{eqnarray}}

\newcommand{\pa}{\partial}
\renewcommand{\a}{\alpha}

\renewcommand{\b}{\beta}

\newcommand{\q}{\theta}

\newcommand{\be}{\begin{equation}}
\newcommand{\ee}{\end{equation}}

\begin{document}
\immediate\write16{<<WARNING: LINEDRAW macros work with emTeX-dvivers
                    and other drivers supporting emTeX \special's
                    (dviscr, dvihplj, dvidot, dvips, dviwin, etc.) >>}

\newdimen\Lengthunit       \Lengthunit  = 1.5cm
\newcount\Nhalfperiods     \Nhalfperiods= 9
\newcount\magnitude        \magnitude = 1000

\catcode`\*=11
\newdimen\L*   \newdimen\d*   \newdimen\d**
\newdimen\dm*  \newdimen\dd*  \newdimen\dt*
\newdimen\a*   \newdimen\b*   \newdimen\c*
\newdimen\a**  \newdimen\b**
\newdimen\xL*  \newdimen\yL*
\newdimen\rx*  \newdimen\ry*
\newdimen\tmp* \newdimen\linwid*

\newcount\k*   \newcount\l*   \newcount\m*
\newcount\k**  \newcount\l**  \newcount\m**
\newcount\n*   \newcount\dn*  \newcount\r*
\newcount\N*   \newcount\*one \newcount\*two  \*one=1 \*two=2
\newcount\*ths \*ths=1000
\newcount\angle*  \newcount\q*  \newcount\q**
\newcount\angle** \angle**=0
\newcount\sc*     \sc*=0

\newtoks\cos*  \cos*={1}
\newtoks\sin*  \sin*={0}

\catcode`\[=13

\def\rotate(#1){\advance\angle**#1\angle*=\angle**
\q**=\angle*\ifnum\q**<0\q**=-\q**\fi
\ifnum\q**>360\q*=\angle*\divide\q*360\multiply\q*360\advance\angle*-\q*\fi
\ifnum\angle*<0\advance\angle*360\fi\q**=\angle*\divide\q**90\q**=\q**
\def\sgcos*{+}\def\sgsin*{+}\relax
\ifcase\q**\or
 \def\sgcos*{-}\def\sgsin*{+}\or
 \def\sgcos*{-}\def\sgsin*{-}\or
 \def\sgcos*{+}\def\sgsin*{-}\else\fi
\q*=\q**
\multiply\q*90\advance\angle*-\q*
\ifnum\angle*>45\sc*=1\angle*=-\angle*\advance\angle*90\else\sc*=0\fi
\def[##1,##2]{\ifnum\sc*=0\relax
\edef\cs*{\sgcos*.##1}\edef\sn*{\sgsin*.##2}\ifcase\q**\or
 \edef\cs*{\sgcos*.##2}\edef\sn*{\sgsin*.##1}\or
 \edef\cs*{\sgcos*.##1}\edef\sn*{\sgsin*.##2}\or
 \edef\cs*{\sgcos*.##2}\edef\sn*{\sgsin*.##1}\else\fi\else
\edef\cs*{\sgcos*.##2}\edef\sn*{\sgsin*.##1}\ifcase\q**\or
 \edef\cs*{\sgcos*.##1}\edef\sn*{\sgsin*.##2}\or
 \edef\cs*{\sgcos*.##2}\edef\sn*{\sgsin*.##1}\or
 \edef\cs*{\sgcos*.##1}\edef\sn*{\sgsin*.##2}\else\fi\fi
\cos*={\cs*}\sin*={\sn*}\global\edef\gcos*{\cs*}\global\edef\gsin*{\sn*}}\relax
\ifcase\angle*[9999,0]\or
[999,017]\or[999,034]\or[998,052]\or[997,069]\or[996,087]\or
[994,104]\or[992,121]\or[990,139]\or[987,156]\or[984,173]\or
[981,190]\or[978,207]\or[974,224]\or[970,241]\or[965,258]\or
[961,275]\or[956,292]\or[951,309]\or[945,325]\or[939,342]\or
[933,358]\or[927,374]\or[920,390]\or[913,406]\or[906,422]\or
[898,438]\or[891,453]\or[882,469]\or[874,484]\or[866,499]\or
[857,515]\or[848,529]\or[838,544]\or[829,559]\or[819,573]\or
[809,587]\or[798,601]\or[788,615]\or[777,629]\or[766,642]\or
[754,656]\or[743,669]\or[731,681]\or[719,694]\or[707,707]\or
\else[9999,0]\fi}

\catcode`\[=12

\def\GRAPH(hsize=#1)#2{\hbox to #1\Lengthunit{#2\hss}}

\def\Linewidth#1{\global\linwid*=#1\relax
\global\divide\linwid*10\global\multiply\linwid*\mag
\global\divide\linwid*100\special{em:linewidth \the\linwid*}}

\Linewidth{.4pt}
\def\sm*{\special{em:moveto}}
\def\sl*{\special{em:lineto}}
\let\moveto=\sm*
\let\lineto=\sl*
\newbox\spm*   \newbox\spl*
\setbox\spm*\hbox{\sm*}
\setbox\spl*\hbox{\sl*}

\def\mov#1(#2,#3)#4{\rlap{\L*=#1\Lengthunit
\xL*=#2\L* \yL*=#3\L*
\xL*=\xscale\xL* \yL*=\yscale\yL*
\rx* \the\cos*\xL* \tmp* \the\sin*\yL* \advance\rx*-\tmp*
\ry* \the\cos*\yL* \tmp* \the\sin*\xL* \advance\ry*\tmp*
\kern\rx*\raise\ry*\hbox{#4}}}

\def\rmov*(#1,#2)#3{\rlap{\xL*=#1\yL*=#2\relax
\rx* \the\cos*\xL* \tmp* \the\sin*\yL* \advance\rx*-\tmp*
\ry* \the\cos*\yL* \tmp* \the\sin*\xL* \advance\ry*\tmp*
\kern\rx*\raise\ry*\hbox{#3}}}

\def\lin#1(#2,#3){\rlap{\sm*\mov#1(#2,#3){\sl*}}}

\def\arr*(#1,#2,#3){\rmov*(#1\dd*,#1\dt*){\sm*
\rmov*(#2\dd*,#2\dt*){\rmov*(#3\dt*,-#3\dd*){\sl*}}\sm*
\rmov*(#2\dd*,#2\dt*){\rmov*(-#3\dt*,#3\dd*){\sl*}}}}

\def\arrow#1(#2,#3){\rlap{\lin#1(#2,#3)\mov#1(#2,#3){\relax
\d**=-.012\Lengthunit\dd*=#2\d**\dt*=#3\d**
\arr*(1,10,4)\arr*(3,8,4)\arr*(4.8,4.2,3)}}}

\def\arrlin#1(#2,#3){\rlap{\L*=#1\Lengthunit\L*=.5\L*
\lin#1(#2,#3)\rmov*(#2\L*,#3\L*){\arrow.1(#2,#3)}}}

\def\dasharrow#1(#2,#3){\rlap{{\Lengthunit=0.9\Lengthunit
\dashlin#1(#2,#3)\mov#1(#2,#3){\sm*}}\mov#1(#2,#3){\sl*
\d**=-.012\Lengthunit\dd*=#2\d**\dt*=#3\d**
\arr*(1,10,4)\arr*(3,8,4)\arr*(4.8,4.2,3)}}}

\def\clap#1{\hbox to 0pt{\hss #1\hss}}

\def\ind(#1,#2)#3{\rlap{\L*=.1\Lengthunit
\xL*=#1\L* \yL*=#2\L*
\rx* \the\cos*\xL* \tmp* \the\sin*\yL* \advance\rx*-\tmp*
\ry* \the\cos*\yL* \tmp* \the\sin*\xL* \advance\ry*\tmp*
\kern\rx*\raise\ry*\hbox{\lower2pt\clap{$#3$}}}}

\def\sh*(#1,#2)#3{\rlap{\dm*=\the\n*\d**
\xL*=\xscale\dm* \yL*=\yscale\dm* \xL*=#1\xL* \yL*=#2\yL*
\rx* \the\cos*\xL* \tmp* \the\sin*\yL* \advance\rx*-\tmp*
\ry* \the\cos*\yL* \tmp* \the\sin*\xL* \advance\ry*\tmp*
\kern\rx*\raise\ry*\hbox{#3}}}

\def\calcnum*#1(#2,#3){\a*=1000sp\b*=1000sp\a*=#2\a*\b*=#3\b*
\ifdim\a*<0pt\a*-\a*\fi\ifdim\b*<0pt\b*-\b*\fi
\ifdim\a*>\b*\c*=.96\a*\advance\c*.4\b*
\else\c*=.96\b*\advance\c*.4\a*\fi
\k*\a*\multiply\k*\k*\l*\b*\multiply\l*\l*
\m*\k*\advance\m*\l*\n*\c*\r*\n*\multiply\n*\n*
\dn*\m*\advance\dn*-\n*\divide\dn*2\divide\dn*\r*
\advance\r*\dn*
\c*=\the\Nhalfperiods5sp\c*=#1\c*\ifdim\c*<0pt\c*-\c*\fi
\multiply\c*\r*\N*\c*\divide\N*10000}

\def\dashlin#1(#2,#3){\rlap{\calcnum*#1(#2,#3)\relax
\d**=#1\Lengthunit\ifdim\d**<0pt\d**-\d**\fi
\divide\N*2\multiply\N*2\advance\N*\*one
\divide\d**\N*\sm*\n*\*one\sh*(#2,#3){\sl*}\loop
\advance\n*\*one\sh*(#2,#3){\sm*}\advance\n*\*one
\sh*(#2,#3){\sl*}\ifnum\n*<\N*\repeat}}

\def\dashdotlin#1(#2,#3){\rlap{\calcnum*#1(#2,#3)\relax
\d**=#1\Lengthunit\ifdim\d**<0pt\d**-\d**\fi
\divide\N*2\multiply\N*2\advance\N*1\multiply\N*2\relax
\divide\d**\N*\sm*\n*\*two\sh*(#2,#3){\sl*}\loop
\advance\n*\*one\sh*(#2,#3){\kern-1.48pt\lower.5pt\hbox{\rm.}}\relax
\advance\n*\*one\sh*(#2,#3){\sm*}\advance\n*\*two
\sh*(#2,#3){\sl*}\ifnum\n*<\N*\repeat}}

\def\shl*(#1,#2)#3{\kern#1#3\lower#2#3\hbox{\unhcopy\spl*}}

\def\trianglin#1(#2,#3){\rlap{\toks0={#2}\toks1={#3}\calcnum*#1(#2,#3)\relax
\dd*=.57\Lengthunit\dd*=#1\dd*\divide\dd*\N*
\divide\dd*\*ths \multiply\dd*\magnitude
\d**=#1\Lengthunit\ifdim\d**<0pt\d**-\d**\fi
\multiply\N*2\divide\d**\N*\sm*\n*\*one\loop
\shl**{\dd*}\dd*-\dd*\advance\n*2\relax
\ifnum\n*<\N*\repeat\n*\N*\shl**{0pt}}}

\def\wavelin#1(#2,#3){\rlap{\toks0={#2}\toks1={#3}\calcnum*#1(#2,#3)\relax
\dd*=.23\Lengthunit\dd*=#1\dd*\divide\dd*\N*
\divide\dd*\*ths \multiply\dd*\magnitude
\d**=#1\Lengthunit\ifdim\d**<0pt\d**-\d**\fi
\multiply\N*4\divide\d**\N*\sm*\n*\*one\loop
\shl**{\dd*}\dt*=1.3\dd*\advance\n*\*one
\shl**{\dt*}\advance\n*\*one
\shl**{\dd*}\advance\n*\*two
\dd*-\dd*\ifnum\n*<\N*\repeat\n*\N*\shl**{0pt}}}

\def\w*lin(#1,#2){\rlap{\toks0={#1}\toks1={#2}\d**=\Lengthunit\dd*=-.12\d**
\divide\dd*\*ths \multiply\dd*\magnitude
\N*8\divide\d**\N*\sm*\n*\*one\loop
\shl**{\dd*}\dt*=1.3\dd*\advance\n*\*one
\shl**{\dt*}\advance\n*\*one
\shl**{\dd*}\advance\n*\*one
\shl**{0pt}\dd*-\dd*\advance\n*1\ifnum\n*<\N*\repeat}}

\def\l*arc(#1,#2)[#3][#4]{\rlap{\toks0={#1}\toks1={#2}\d**=\Lengthunit
\dd*=#3.037\d**\dd*=#4\dd*\dt*=#3.049\d**\dt*=#4\dt*\ifdim\d**>10mm\relax
\d**=.25\d**\n*\*one\shl**{-\dd*}\n*\*two\shl**{-\dt*}\n*3\relax
\shl**{-\dd*}\n*4\relax\shl**{0pt}\else
\ifdim\d**>5mm\d**=.5\d**\n*\*one\shl**{-\dt*}\n*\*two
\shl**{0pt}\else\n*\*one\shl**{0pt}\fi\fi}}

\def\d*arc(#1,#2)[#3][#4]{\rlap{\toks0={#1}\toks1={#2}\d**=\Lengthunit
\dd*=#3.037\d**\dd*=#4\dd*\d**=.25\d**\sm*\n*\*one\shl**{-\dd*}\relax
\n*3\relax\sh*(#1,#2){\xL*=\xscale\dd*\yL*=\yscale\dd*
\kern#2\xL*\lower#1\yL*\hbox{\sm*}}\n*4\relax\shl**{0pt}}}

\def\shl**#1{\c*=\the\n*\d**\d*=#1\relax
\a*=\the\toks0\c*\b*=\the\toks1\d*\advance\a*-\b*
\b*=\the\toks1\c*\d*=\the\toks0\d*\advance\b*\d*
\a*=\xscale\a*\b*=\yscale\b*
\rx* \the\cos*\a* \tmp* \the\sin*\b* \advance\rx*-\tmp*
\ry* \the\cos*\b* \tmp* \the\sin*\a* \advance\ry*\tmp*
\raise\ry*\rlap{\kern\rx*\unhcopy\spl*}}

\def\wlin*#1(#2,#3)[#4]{\rlap{\toks0={#2}\toks1={#3}\relax
\c*=#1\l*\c*\c*=.01\Lengthunit\m*\c*\divide\l*\m*
\c*=\the\Nhalfperiods5sp\multiply\c*\l*\N*\c*\divide\N*\*ths
\divide\N*2\multiply\N*2\advance\N*\*one
\dd*=.002\Lengthunit\dd*=#4\dd*\multiply\dd*\l*\divide\dd*\N*
\divide\dd*\*ths \multiply\dd*\magnitude
\d**=#1\multiply\N*4\divide\d**\N*\sm*\n*\*one\loop
\shl**{\dd*}\dt*=1.3\dd*\advance\n*\*one
\shl**{\dt*}\advance\n*\*one
\shl**{\dd*}\advance\n*\*two
\dd*-\dd*\ifnum\n*<\N*\repeat\n*\N*\shl**{0pt}}}

\def\wavebox#1{\setbox0\hbox{#1}\relax
\a*=\wd0\advance\a*14pt\b*=\ht0\advance\b*\dp0\advance\b*14pt\relax
\hbox{\kern9pt\relax
\rmov*(0pt,\ht0){\rmov*(-7pt,7pt){\wlin*\a*(1,0)[+]\wlin*\b*(0,-1)[-]}}\relax
\rmov*(\wd0,-\dp0){\rmov*(7pt,-7pt){\wlin*\a*(-1,0)[+]\wlin*\b*(0,1)[-]}}\relax
\box0\kern9pt}}

\def\rectangle#1(#2,#3){\relax
\lin#1(#2,0)\lin#1(0,#3)\mov#1(0,#3){\lin#1(#2,0)}\mov#1(#2,0){\lin#1(0,#3)}}

\def\dashrectangle#1(#2,#3){\dashlin#1(#2,0)\dashlin#1(0,#3)\relax
\mov#1(0,#3){\dashlin#1(#2,0)}\mov#1(#2,0){\dashlin#1(0,#3)}}

\def\waverectangle#1(#2,#3){\L*=#1\Lengthunit\a*=#2\L*\b*=#3\L*
\ifdim\a*<0pt\a*-\a*\def\x*{-1}\else\def\x*{1}\fi
\ifdim\b*<0pt\b*-\b*\def\y*{-1}\else\def\y*{1}\fi
\wlin*\a*(\x*,0)[-]\wlin*\b*(0,\y*)[+]\relax
\mov#1(0,#3){\wlin*\a*(\x*,0)[+]}\mov#1(#2,0){\wlin*\b*(0,\y*)[-]}}

\def\calcparab*{\ifnum\n*>\m*\k*\N*\advance\k*-\n*\else\k*\n*\fi
\a*=\the\k* sp\a*=10\a*\b*\dm*\advance\b*-\a*\k*\b*
\a*=\the\*ths\b*\divide\a*\l*\multiply\a*\k*
\divide\a*\l*\k*\*ths\r*\a*\advance\k*-\r*\dt*=\the\k*\L*}

\def\arcto#1(#2,#3)[#4]{\rlap{\toks0={#2}\toks1={#3}\calcnum*#1(#2,#3)\relax
\dm*=135sp\dm*=#1\dm*\d**=#1\Lengthunit\ifdim\dm*<0pt\dm*-\dm*\fi
\multiply\dm*\r*\a*=.3\dm*\a*=#4\a*\ifdim\a*<0pt\a*-\a*\fi
\advance\dm*\a*\N*\dm*\divide\N*10000\relax
\divide\N*2\multiply\N*2\advance\N*\*one
\L*=-.25\d**\L*=#4\L*\divide\d**\N*\divide\L*\*ths
\m*\N*\divide\m*2\dm*=\the\m*5sp\l*\dm*\sm*\n*\*one\loop
\calcparab*\shl**{-\dt*}\advance\n*1\ifnum\n*<\N*\repeat}}

\def\arrarcto#1(#2,#3)[#4]{\L*=#1\Lengthunit\L*=.54\L*
\arcto#1(#2,#3)[#4]\rmov*(#2\L*,#3\L*){\d*=.457\L*\d*=#4\d*\d**-\d*
\rmov*(#3\d**,#2\d*){\arrow.02(#2,#3)}}}

\def\dasharcto#1(#2,#3)[#4]{\rlap{\toks0={#2}\toks1={#3}\relax
\calcnum*#1(#2,#3)\dm*=\the\N*5sp\a*=.3\dm*\a*=#4\a*\ifdim\a*<0pt\a*-\a*\fi
\advance\dm*\a*\N*\dm*
\divide\N*20\multiply\N*2\advance\N*1\d**=#1\Lengthunit
\L*=-.25\d**\L*=#4\L*\divide\d**\N*\divide\L*\*ths
\m*\N*\divide\m*2\dm*=\the\m*5sp\l*\dm*
\sm*\n*\*one\loop\calcparab*
\shl**{-\dt*}\advance\n*1\ifnum\n*>\N*\else\calcparab*
\sh*(#2,#3){\xL*=#3\dt* \yL*=#2\dt*
\rx* \the\cos*\xL* \tmp* \the\sin*\yL* \advance\rx*\tmp*
\ry* \the\cos*\yL* \tmp* \the\sin*\xL* \advance\ry*-\tmp*
\kern\rx*\lower\ry*\hbox{\sm*}}\fi
\advance\n*1\ifnum\n*<\N*\repeat}}

\def\*shl*#1{\c*=\the\n*\d**\advance\c*#1\a**\d*\dt*\advance\d*#1\b**
\a*=\the\toks0\c*\b*=\the\toks1\d*\advance\a*-\b*
\b*=\the\toks1\c*\d*=\the\toks0\d*\advance\b*\d*
\rx* \the\cos*\a* \tmp* \the\sin*\b* \advance\rx*-\tmp*
\ry* \the\cos*\b* \tmp* \the\sin*\a* \advance\ry*\tmp*
\raise\ry*\rlap{\kern\rx*\unhcopy\spl*}}

\def\calcnormal*#1{\b**=10000sp\a**\b**\k*\n*\advance\k*-\m*
\multiply\a**\k*\divide\a**\m*\a**=#1\a**\ifdim\a**<0pt\a**-\a**\fi
\ifdim\a**>\b**\d*=.96\a**\advance\d*.4\b**
\else\d*=.96\b**\advance\d*.4\a**\fi
\d*=.01\d*\r*\d*\divide\a**\r*\divide\b**\r*
\ifnum\k*<0\a**-\a**\fi\d*=#1\d*\ifdim\d*<0pt\b**-\b**\fi
\k*\a**\a**=\the\k*\dd*\k*\b**\b**=\the\k*\dd*}

\def\wavearcto#1(#2,#3)[#4]{\rlap{\toks0={#2}\toks1={#3}\relax
\calcnum*#1(#2,#3)\c*=\the\N*5sp\a*=.4\c*\a*=#4\a*\ifdim\a*<0pt\a*-\a*\fi
\advance\c*\a*\N*\c*\divide\N*20\multiply\N*2\advance\N*-1\multiply\N*4\relax
\d**=#1\Lengthunit\dd*=.012\d**
\divide\dd*\*ths \multiply\dd*\magnitude
\ifdim\d**<0pt\d**-\d**\fi\L*=.25\d**
\divide\d**\N*\divide\dd*\N*\L*=#4\L*\divide\L*\*ths
\m*\N*\divide\m*2\dm*=\the\m*0sp\l*\dm*
\sm*\n*\*one\loop\calcnormal*{#4}\calcparab*
\*shl*{1}\advance\n*\*one\calcparab*
\*shl*{1.3}\advance\n*\*one\calcparab*
\*shl*{1}\advance\n*2\dd*-\dd*\ifnum\n*<\N*\repeat\n*\N*\shl**{0pt}}}

\def\triangarcto#1(#2,#3)[#4]{\rlap{\toks0={#2}\toks1={#3}\relax
\calcnum*#1(#2,#3)\c*=\the\N*5sp\a*=.4\c*\a*=#4\a*\ifdim\a*<0pt\a*-\a*\fi
\advance\c*\a*\N*\c*\divide\N*20\multiply\N*2\advance\N*-1\multiply\N*2\relax
\d**=#1\Lengthunit\dd*=.012\d**
\divide\dd*\*ths \multiply\dd*\magnitude
\ifdim\d**<0pt\d**-\d**\fi\L*=.25\d**
\divide\d**\N*\divide\dd*\N*\L*=#4\L*\divide\L*\*ths
\m*\N*\divide\m*2\dm*=\the\m*0sp\l*\dm*
\sm*\n*\*one\loop\calcnormal*{#4}\calcparab*
\*shl*{1}\advance\n*2\dd*-\dd*\ifnum\n*<\N*\repeat\n*\N*\shl**{0pt}}}

\def\hr*#1{\L*=\xscale\Lengthunit\ifnum
\angle**=0\clap{\vrule width#1\L* height.1pt}\else
\L*=#1\L*\L*=.5\L*\rmov*(-\L*,0pt){\sm*}\rmov*(\L*,0pt){\sl*}\fi}

\def\shade#1[#2]{\rlap{\Lengthunit=#1\Lengthunit
\special{em:linewidth .001pt}\relax
\mov(0,#2.05){\hr*{.994}}\mov(0,#2.1){\hr*{.980}}\relax
\mov(0,#2.15){\hr*{.953}}\mov(0,#2.2){\hr*{.916}}\relax
\mov(0,#2.25){\hr*{.867}}\mov(0,#2.3){\hr*{.798}}\relax
\mov(0,#2.35){\hr*{.715}}\mov(0,#2.4){\hr*{.603}}\relax
\mov(0,#2.45){\hr*{.435}}\special{em:linewidth \the\linwid*}}}

\def\dshade#1[#2]{\rlap{\special{em:linewidth .001pt}\relax
\Lengthunit=#1\Lengthunit\if#2-\def\t*{+}\else\def\t*{-}\fi
\mov(0,\t*.025){\relax
\mov(0,#2.05){\hr*{.995}}\mov(0,#2.1){\hr*{.988}}\relax
\mov(0,#2.15){\hr*{.969}}\mov(0,#2.2){\hr*{.937}}\relax
\mov(0,#2.25){\hr*{.893}}\mov(0,#2.3){\hr*{.836}}\relax
\mov(0,#2.35){\hr*{.760}}\mov(0,#2.4){\hr*{.662}}\relax
\mov(0,#2.45){\hr*{.531}}\mov(0,#2.5){\hr*{.320}}\relax
\special{em:linewidth \the\linwid*}}}}

\def\vdot{\rlap{\kern-1.9pt\lower1.8pt\hbox{$\scriptstyle\bullet$}}}
\def\vtimes{\rlap{\kern-3pt\lower1.8pt\hbox{$\scriptstyle\times$}}}
\def\vDot{\rlap{\kern-2.3pt\lower2.7pt\hbox{$\bullet$}}}
\def\vTimes{\rlap{\kern-3.6pt\lower2.4pt\hbox{$\times$}}}

\def\arc(#1)[#2,#3]{{\k*=#2\l*=#3\m*=\l*
\advance\m*-6\ifnum\k*>\l*\relax\else
{\rotate(#2)\mov(#1,0){\sm*}}\loop
\ifnum\k*<\m*\advance\k*5{\rotate(\k*)\mov(#1,0){\sl*}}\repeat
{\rotate(#3)\mov(#1,0){\sl*}}\fi}}

\def\dasharc(#1)[#2,#3]{{\k**=#2\n*=#3\advance\n*-1\advance\n*-\k**
\L*=1000sp\L*#1\L* \multiply\L*\n* \multiply\L*\Nhalfperiods
\divide\L*57\N*\L* \divide\N*2000\ifnum\N*=0\N*1\fi
\r*\n*  \divide\r*\N* \ifnum\r*<2\r*2\fi
\m**\r* \divide\m**2 \l**\r* \advance\l**-\m** \N*\n* \divide\N*\r*
\k**\r* \multiply\k**\N* \dn*\n*
\advance\dn*-\k** \divide\dn*2\advance\dn*\*one
\r*\l** \divide\r*2\advance\dn*\r* \advance\N*-2\k**#2\relax
\ifnum\l**<6{\rotate(#2)\mov(#1,0){\sm*}}\advance\k**\dn*
{\rotate(\k**)\mov(#1,0){\sl*}}\advance\k**\m**
{\rotate(\k**)\mov(#1,0){\sm*}}\loop
\advance\k**\l**{\rotate(\k**)\mov(#1,0){\sl*}}\advance\k**\m**
{\rotate(\k**)\mov(#1,0){\sm*}}\advance\N*-1\ifnum\N*>0\repeat
{\rotate(#3)\mov(#1,0){\sl*}}\else\advance\k**\dn*
\arc(#1)[#2,\k**]\loop\advance\k**\m** \r*\k**
\advance\k**\l** {\arc(#1)[\r*,\k**]}\relax
\advance\N*-1\ifnum\N*>0\repeat
\advance\k**\m**\arc(#1)[\k**,#3]\fi}}

\def\triangarc#1(#2)[#3,#4]{{\k**=#3\n*=#4\advance\n*-\k**
\L*=1000sp\L*#2\L* \multiply\L*\n* \multiply\L*\Nhalfperiods
\divide\L*57\N*\L* \divide\N*1000\ifnum\N*=0\N*1\fi
\d**=#2\Lengthunit \d*\d** \divide\d*57\multiply\d*\n*
\r*\n*  \divide\r*\N* \ifnum\r*<2\r*2\fi
\m**\r* \divide\m**2 \l**\r* \advance\l**-\m** \N*\n* \divide\N*\r*
\dt*\d* \divide\dt*\N* \dt*.5\dt* \dt*#1\dt*
\divide\dt*1000\multiply\dt*\magnitude
\k**\r* \multiply\k**\N* \dn*\n* \advance\dn*-\k** \divide\dn*2\relax
\r*\l** \divide\r*2\advance\dn*\r* \advance\N*-1\k**#3\relax
{\rotate(#3)\mov(#2,0){\sm*}}\advance\k**\dn*
{\rotate(\k**)\mov(#2,0){\sl*}}\advance\k**-\m**\advance\l**\m**\loop\dt*-\dt*
\d*\d** \advance\d*\dt*
\advance\k**\l**{\rotate(\k**)\rmov*(\d*,0pt){\sl*}}%
\advance\N*-1\ifnum\N*>0\repeat\advance\k**\m**
{\rotate(\k**)\mov(#2,0){\sl*}}{\rotate(#4)\mov(#2,0){\sl*}}}}

\def\wavearc#1(#2)[#3,#4]{{\k**=#3\n*=#4\advance\n*-\k**
\L*=4000sp\L*#2\L* \multiply\L*\n* \multiply\L*\Nhalfperiods
\divide\L*57\N*\L* \divide\N*1000\ifnum\N*=0\N*1\fi
\d**=#2\Lengthunit \d*\d** \divide\d*57\multiply\d*\n*
\r*\n*  \divide\r*\N* \ifnum\r*=0\r*1\fi
\m**\r* \divide\m**2 \l**\r* \advance\l**-\m** \N*\n* \divide\N*\r*
\dt*\d* \divide\dt*\N* \dt*.7\dt* \dt*#1\dt*
\divide\dt*1000\multiply\dt*\magnitude
\k**\r* \multiply\k**\N* \dn*\n* \advance\dn*-\k** \divide\dn*2\relax
\divide\N*4\advance\N*-1\k**#3\relax
{\rotate(#3)\mov(#2,0){\sm*}}\advance\k**\dn*
{\rotate(\k**)\mov(#2,0){\sl*}}\advance\k**-\m**\advance\l**\m**\loop\dt*-\dt*
\d*\d** \advance\d*\dt* \dd*\d** \advance\dd*1.3\dt*
\advance\k**\r*{\rotate(\k**)\rmov*(\d*,0pt){\sl*}}\relax
\advance\k**\r*{\rotate(\k**)\rmov*(\dd*,0pt){\sl*}}\relax
\advance\k**\r*{\rotate(\k**)\rmov*(\d*,0pt){\sl*}}\relax
\advance\k**\r*
\advance\N*-1\ifnum\N*>0\repeat\advance\k**\m**
{\rotate(\k**)\mov(#2,0){\sl*}}{\rotate(#4)\mov(#2,0){\sl*}}}}

\def\gmov*#1(#2,#3)#4{\rlap{\L*=#1\Lengthunit
\xL*=#2\L* \yL*=#3\L*
\rx* \gcos*\xL* \tmp* \gsin*\yL* \advance\rx*-\tmp*
\ry* \gcos*\yL* \tmp* \gsin*\xL* \advance\ry*\tmp*
\rx*=\xscale\rx* \ry*=\yscale\ry*
\xL* \the\cos*\rx* \tmp* \the\sin*\ry* \advance\xL*-\tmp*
\yL* \the\cos*\ry* \tmp* \the\sin*\rx* \advance\yL*\tmp*
\kern\xL*\raise\yL*\hbox{#4}}}

\def\rgmov*(#1,#2)#3{\rlap{\xL*#1\yL*#2\relax
\rx* \gcos*\xL* \tmp* \gsin*\yL* \advance\rx*-\tmp*
\ry* \gcos*\yL* \tmp* \gsin*\xL* \advance\ry*\tmp*
\rx*=\xscale\rx* \ry*=\yscale\ry*
\xL* \the\cos*\rx* \tmp* \the\sin*\ry* \advance\xL*-\tmp*
\yL* \the\cos*\ry* \tmp* \the\sin*\rx* \advance\yL*\tmp*
\kern\xL*\raise\yL*\hbox{#3}}}

\def\Earc(#1)[#2,#3][#4,#5]{{\k*=#2\l*=#3\m*=\l*
\advance\m*-6\ifnum\k*>\l*\relax\else\def\xscale{#4}\def\yscale{#5}\relax
{\angle**0\rotate(#2)}\gmov*(#1,0){\sm*}\loop
\ifnum\k*<\m*\advance\k*5\relax
{\angle**0\rotate(\k*)}\gmov*(#1,0){\sl*}\repeat
{\angle**0\rotate(#3)}\gmov*(#1,0){\sl*}\relax
\def\xscale{1}\def\yscale{1}\fi}}

\def\dashEarc(#1)[#2,#3][#4,#5]{{\k**=#2\n*=#3\advance\n*-1\advance\n*-\k**
\L*=1000sp\L*#1\L* \multiply\L*\n* \multiply\L*\Nhalfperiods
\divide\L*57\N*\L* \divide\N*2000\ifnum\N*=0\N*1\fi
\r*\n*  \divide\r*\N* \ifnum\r*<2\r*2\fi
\m**\r* \divide\m**2 \l**\r* \advance\l**-\m** \N*\n* \divide\N*\r*
\k**\r*\multiply\k**\N* \dn*\n* \advance\dn*-\k** \divide\dn*2\advance\dn*\*one
\r*\l** \divide\r*2\advance\dn*\r* \advance\N*-2\k**#2\relax
\ifnum\l**<6\def\xscale{#4}\def\yscale{#5}\relax
{\angle**0\rotate(#2)}\gmov*(#1,0){\sm*}\advance\k**\dn*
{\angle**0\rotate(\k**)}\gmov*(#1,0){\sl*}\advance\k**\m**
{\angle**0\rotate(\k**)}\gmov*(#1,0){\sm*}\loop
\advance\k**\l**{\angle**0\rotate(\k**)}\gmov*(#1,0){\sl*}\advance\k**\m**
{\angle**0\rotate(\k**)}\gmov*(#1,0){\sm*}\advance\N*-1\ifnum\N*>0\repeat
{\angle**0\rotate(#3)}\gmov*(#1,0){\sl*}\def\xscale{1}\def\yscale{1}\else
\advance\k**\dn* \Earc(#1)[#2,\k**][#4,#5]\loop\advance\k**\m** \r*\k**
\advance\k**\l** {\Earc(#1)[\r*,\k**][#4,#5]}\relax
\advance\N*-1\ifnum\N*>0\repeat
\advance\k**\m**\Earc(#1)[\k**,#3][#4,#5]\fi}}

\def\triangEarc#1(#2)[#3,#4][#5,#6]{{\k**=#3\n*=#4\advance\n*-\k**
\L*=1000sp\L*#2\L* \multiply\L*\n* \multiply\L*\Nhalfperiods
\divide\L*57\N*\L* \divide\N*1000\ifnum\N*=0\N*1\fi
\d**=#2\Lengthunit \d*\d** \divide\d*57\multiply\d*\n*
\r*\n*  \divide\r*\N* \ifnum\r*<2\r*2\fi
\m**\r* \divide\m**2 \l**\r* \advance\l**-\m** \N*\n* \divide\N*\r*
\dt*\d* \divide\dt*\N* \dt*.5\dt* \dt*#1\dt*
\divide\dt*1000\multiply\dt*\magnitude
\k**\r* \multiply\k**\N* \dn*\n* \advance\dn*-\k** \divide\dn*2\relax
\r*\l** \divide\r*2\advance\dn*\r* \advance\N*-1\k**#3\relax
\def\xscale{#5}\def\yscale{#6}\relax
{\angle**0\rotate(#3)}\gmov*(#2,0){\sm*}\advance\k**\dn*
{\angle**0\rotate(\k**)}\gmov*(#2,0){\sl*}\advance\k**-\m**
\advance\l**\m**\loop\dt*-\dt* \d*\d** \advance\d*\dt*
\advance\k**\l**{\angle**0\rotate(\k**)}\rgmov*(\d*,0pt){\sl*}\relax
\advance\N*-1\ifnum\N*>0\repeat\advance\k**\m**
{\angle**0\rotate(\k**)}\gmov*(#2,0){\sl*}\relax
{\angle**0\rotate(#4)}\gmov*(#2,0){\sl*}\def\xscale{1}\def\yscale{1}}}

\def\waveEarc#1(#2)[#3,#4][#5,#6]{{\k**=#3\n*=#4\advance\n*-\k**
\L*=4000sp\L*#2\L* \multiply\L*\n* \multiply\L*\Nhalfperiods
\divide\L*57\N*\L* \divide\N*1000\ifnum\N*=0\N*1\fi
\d**=#2\Lengthunit \d*\d** \divide\d*57\multiply\d*\n*
\r*\n*  \divide\r*\N* \ifnum\r*=0\r*1\fi
\m**\r* \divide\m**2 \l**\r* \advance\l**-\m** \N*\n* \divide\N*\r*
\dt*\d* \divide\dt*\N* \dt*.7\dt* \dt*#1\dt*
\divide\dt*1000\multiply\dt*\magnitude
\k**\r* \multiply\k**\N* \dn*\n* \advance\dn*-\k** \divide\dn*2\relax
\divide\N*4\advance\N*-1\k**#3\def\xscale{#5}\def\yscale{#6}\relax
{\angle**0\rotate(#3)}\gmov*(#2,0){\sm*}\advance\k**\dn*
{\angle**0\rotate(\k**)}\gmov*(#2,0){\sl*}\advance\k**-\m**
\advance\l**\m**\loop\dt*-\dt*
\d*\d** \advance\d*\dt* \dd*\d** \advance\dd*1.3\dt*
\advance\k**\r*{\angle**0\rotate(\k**)}\rgmov*(\d*,0pt){\sl*}\relax
\advance\k**\r*{\angle**0\rotate(\k**)}\rgmov*(\dd*,0pt){\sl*}\relax
\advance\k**\r*{\angle**0\rotate(\k**)}\rgmov*(\d*,0pt){\sl*}\relax
\advance\k**\r*
\advance\N*-1\ifnum\N*>0\repeat\advance\k**\m**
{\angle**0\rotate(\k**)}\gmov*(#2,0){\sl*}\relax
{\angle**0\rotate(#4)}\gmov*(#2,0){\sl*}\def\xscale{1}\def\yscale{1}}}

\newcount\CatcodeOfAtSign
\CatcodeOfAtSign=\the\catcode`\@
\catcode`\@=11
\def\@arc#1[#2][#3]{\rlap{\Lengthunit=#1\Lengthunit
\sm*\l*arc(#2.1914,#3.0381)[#2][#3]\relax
\mov(#2.1914,#3.0381){\l*arc(#2.1622,#3.1084)[#2][#3]}\relax
\mov(#2.3536,#3.1465){\l*arc(#2.1084,#3.1622)[#2][#3]}\relax
\mov(#2.4619,#3.3086){\l*arc(#2.0381,#3.1914)[#2][#3]}}}

\def\dash@arc#1[#2][#3]{\rlap{\Lengthunit=#1\Lengthunit
\d*arc(#2.1914,#3.0381)[#2][#3]\relax
\mov(#2.1914,#3.0381){\d*arc(#2.1622,#3.1084)[#2][#3]}\relax
\mov(#2.3536,#3.1465){\d*arc(#2.1084,#3.1622)[#2][#3]}\relax
\mov(#2.4619,#3.3086){\d*arc(#2.0381,#3.1914)[#2][#3]}}}

\def\wave@arc#1[#2][#3]{\rlap{\Lengthunit=#1\Lengthunit
\w*lin(#2.1914,#3.0381)\relax
\mov(#2.1914,#3.0381){\w*lin(#2.1622,#3.1084)}\relax
\mov(#2.3536,#3.1465){\w*lin(#2.1084,#3.1622)}\relax
\mov(#2.4619,#3.3086){\w*lin(#2.0381,#3.1914)}}}

\def\bezier#1(#2,#3)(#4,#5)(#6,#7){\N*#1\l*\N* \advance\l*\*one
\d* #4\Lengthunit \advance\d* -#2\Lengthunit \multiply\d* \*two
\b* #6\Lengthunit \advance\b* -#2\Lengthunit
\advance\b*-\d* \divide\b*\N*
\d** #5\Lengthunit \advance\d** -#3\Lengthunit \multiply\d** \*two
\b** #7\Lengthunit \advance\b** -#3\Lengthunit
\advance\b** -\d** \divide\b**\N*
\mov(#2,#3){\sm*{\loop\ifnum\m*<\l*
\a*\m*\b* \advance\a*\d* \divide\a*\N* \multiply\a*\m*
\a**\m*\b** \advance\a**\d** \divide\a**\N* \multiply\a**\m*
\rmov*(\a*,\a**){\unhcopy\spl*}\advance\m*\*one\repeat}}}

\catcode`\*=12

\newcount\n@ast

\def\n@ast@#1{\n@ast0\relax\get@ast@#1\end}
\def\get@ast@#1{\ifx#1\end\let\next\relax\else
\ifx#1*\advance\n@ast1\fi\let\next\get@ast@\fi\next}

\newif\if@up \newif\if@dwn
\def\up@down@#1{\@upfalse\@dwnfalse
\if#1u\@uptrue\fi\if#1U\@uptrue\fi\if#1+\@uptrue\fi
\if#1d\@dwntrue\fi\if#1D\@dwntrue\fi\if#1-\@dwntrue\fi}

\def\halfcirc#1(#2)[#3]{{\Lengthunit=#2\Lengthunit\up@down@{#3}\relax
\if@up\mov(0,.5){\@arc[-][-]\@arc[+][-]}\fi
\if@dwn\mov(0,-.5){\@arc[-][+]\@arc[+][+]}\fi
\def\lft{\mov(0,.5){\@arc[-][-]}\mov(0,-.5){\@arc[-][+]}}\relax
\def\rght{\mov(0,.5){\@arc[+][-]}\mov(0,-.5){\@arc[+][+]}}\relax
\if#3l\lft\fi\if#3L\lft\fi\if#3r\rght\fi\if#3R\rght\fi
\n@ast@{#1}\relax
\ifnum\n@ast>0\if@up\shade[+]\fi\if@dwn\shade[-]\fi\fi
\ifnum\n@ast>1\if@up\dshade[+]\fi\if@dwn\dshade[-]\fi\fi}}

\def\halfdashcirc(#1)[#2]{{\Lengthunit=#1\Lengthunit\up@down@{#2}\relax
\if@up\mov(0,.5){\dash@arc[-][-]\dash@arc[+][-]}\fi
\if@dwn\mov(0,-.5){\dash@arc[-][+]\dash@arc[+][+]}\fi
\def\lft{\mov(0,.5){\dash@arc[-][-]}\mov(0,-.5){\dash@arc[-][+]}}\relax
\def\rght{\mov(0,.5){\dash@arc[+][-]}\mov(0,-.5){\dash@arc[+][+]}}\relax
\if#2l\lft\fi\if#2L\lft\fi\if#2r\rght\fi\if#2R\rght\fi}}

\def\halfwavecirc(#1)[#2]{{\Lengthunit=#1\Lengthunit\up@down@{#2}\relax
\if@up\mov(0,.5){\wave@arc[-][-]\wave@arc[+][-]}\fi
\if@dwn\mov(0,-.5){\wave@arc[-][+]\wave@arc[+][+]}\fi
\def\lft{\mov(0,.5){\wave@arc[-][-]}\mov(0,-.5){\wave@arc[-][+]}}\relax
\def\rght{\mov(0,.5){\wave@arc[+][-]}\mov(0,-.5){\wave@arc[+][+]}}\relax
\if#2l\lft\fi\if#2L\lft\fi\if#2r\rght\fi\if#2R\rght\fi}}

\catcode`\*=11

\def\Circle#1(#2){\halfcirc#1(#2)[u]\halfcirc#1(#2)[d]\n@ast@{#1}\relax
\ifnum\n@ast>0\L*=\xscale\Lengthunit
\ifnum\angle**=0\clap{\vrule width#2\L* height.1pt}\else
\L*=#2\L*\L*=.5\L*\special{em:linewidth .001pt}\relax
\rmov*(-\L*,0pt){\sm*}\rmov*(\L*,0pt){\sl*}\relax
\special{em:linewidth \the\linwid*}\fi\fi}

\catcode`\*=12

\def\wavecirc(#1){\halfwavecirc(#1)[u]\halfwavecirc(#1)[d]}
\def\dashcirc(#1){\halfdashcirc(#1)[u]\halfdashcirc(#1)[d]}

\def\xscale{1}

\def\yscale{1}

\def\Ellipse#1(#2)[#3,#4]{\def\xscale{#3}\def\yscale{#4}\relax
\Circle#1(#2)\def\xscale{1}\def\yscale{1}}

\def\dashEllipse(#1)[#2,#3]{\def\xscale{#2}\def\yscale{#3}\relax
\dashcirc(#1)\def\xscale{1}\def\yscale{1}}

\def\waveEllipse(#1)[#2,#3]{\def\xscale{#2}\def\yscale{#3}\relax
\wavecirc(#1)\def\xscale{1}\def\yscale{1}}

\def\halfEllipse#1(#2)[#3][#4,#5]{\def\xscale{#4}\def\yscale{#5}\relax
\halfcirc#1(#2)[#3]\def\xscale{1}\def\yscale{1}}

\def\halfdashEllipse(#1)[#2][#3,#4]{\def\xscale{#3}\def\yscale{#4}\relax
\halfdashcirc(#1)[#2]\def\xscale{1}\def\yscale{1}}

\def\halfwaveEllipse(#1)[#2][#3,#4]{\def\xscale{#3}\def\yscale{#4}\relax
\halfwavecirc(#1)[#2]\def\xscale{1}\def\yscale{1}}

\catcode`\@=\the\CatcodeOfAtSign

\title{Skyrme-like models and supersymmetry  in 3+1 dimensions}

\author{J. M. Queiruga}
\affiliation{Instituto de F\'\i sica, Universidade de S\~ao Paulo\\
Caixa Postal 66318, 05315-970, S\~ao Paulo, SP, Brazil}
\email{queiruga@if.usp.br}

\begin{abstract}
We construct  supersymmetric extensions  for different Skyrme-like models in 3+1 dimensions. BPS equations and BPS bounds are obtained from  supersymmetry in some cases. We discuss also the emergence of several Skyrme-like models from supersymmetric Yang-Mills theory and Born-Infeld theory in 5 dimensions.
\end{abstract}

\maketitle

\section{Introduction}

The Skyrme model, proposed originally by Skyrme \cite{Skyrme},  is one of the best-known proposals in the study of nonperturbative QCD at low energies. In this model, the physical degrees of freedom corresponding to the pions are encoded in a $SU(2)$ matrix $U$ while baryons emerge as topological solitons called Skyrmions. The original Skyrme model consists of a quadratic and quartic term in derivatives (and optionally a potential term) and it was applied successfully to the study of nuclear matter \cite{ANW}. However, the most general action for the $SU(2)$ pionic fields which is Poincar\'e invariant and possesses the standard Hamiltonian formulation (i.e., is quadratic in time derivatives) has an additional part. Namely, a sextic term, which is proportional to the square of the baryon (topological) current

\be
\mathcal{L}_{0246}=\mathcal{L}_0+\mathcal{L}_2+ \mathcal{L}_4+\mathcal{L}_6,
\ee
where
\be
\mathcal{L}_2=-\lambda_2 \mbox{Tr}\; (R_\mu R^\mu), \;\;\; \mathcal{L}_4=\lambda_4 \mbox{Tr} \; ([R_\mu , R_\nu]^2), \;\;\; \mathcal{L}_6 =-(24\pi^2)^2 \lambda_6 \mathcal{B}_\mu \mathcal{B}^\mu. 
\ee
and $\mathcal{L}_0$ is a nonderivative part i.e., potential. Moreover,
the baryon current is
\be
\mathcal{B}^\mu = \frac{1}{24\pi^2} \epsilon^{\mu \nu \rho \sigma} \mbox{Tr} \; R_\nu R_\rho R_\sigma, 
\ee
and the left invariant current $L_\mu$ is given by
\be
 R_\mu = U^\dagger \partial_\mu U
\ee

 It should be stressed that the sextic term is unavoidable if one would like to apply the Skyrme-like models to dense nuclear matter and neutron stars \cite{stars}, as it provides the leading behavior for the corresponding equation of state at higher densities \cite{therm}. In fact, it has been argued that in the correct Skyrme effective model, at least as the higher nuclei and higher densities are considered, the sextic and the potential should provide a dominant part of the effective action \cite{nearBPS}. This follows from an observation that the original Skyrme proposal leads to too high binding energies and a crystal state of matter. Both effects are in an obvious conflict with the well-known qualitative properties of nuclear matter. On the other hand, the sextic term together with the potential forms a submodel, usually referred to as the BPS Skyrme model, which cures these two issues: it gives zero classical binding energies \cite{bps} and describes a perfect fluid \cite{therm1}, \cite{therm}. Zero binding energies are obviously related to the BPS nature of the BPS Skyrme model, which in consequence lead to other proposals for BPS type generalizations of the Skyrme model: (1) the Sutcliffe model where the BPS limit is obtained by inclusion of infinitely many vector mesons \cite{Sutcliffe} and (2) the Ferreira-Zakrzewski model \cite{FZ}. Also the formulation of the Skyrme model in curved spaces can also lead to the existence of solutions saturating a BPS bound \cite{Canfora}. We remark that there is also a way to arbitrarily reduce the binding energies by adding a kind of ``repulsing" potential into the original Skyrme model \cite{speight}. However, in this proposal, one never gets a proper BPS theory. All this shows that a better understanding of the BPS sectors of the Skyrme type theories is very well motivated by their relevance to nuclear physics.

A little-explored facet of these models is their supersymmetric (SUSY) extension, especially in $3+1$ dimensions (\cite{nepo}, or  \cite{Nitta0} for more general higher derivative models). Supersymmetry provides a natural form of including fermions in these kinds of theories, in addition it can help in the understanding of the BPS structure \cite{Nitta0}. The well-known result of Witten and Olive \cite{Olive} establishes a deep connection between theories with extended supersymmetry, central charges and BPS bounds.  Some progress has been made in lower dimensions \cite{queiruga0}-\cite{bolog}, but several questions still remain open in $3 + 1$ dimensions: the supersymmetric extension of the Skyrme model, or the BPS Skyrme model, and the connection between Bogomolny bounds and central charges of the superalgebra, etc. In this work we will try to answer some of these questions.

Some years ago Sakai and Sugimoto \cite{Sakai}, using a D-brane construction and holographic methods, obtained the Skyrme model coupled to an infinite tower of vector mesons from a Yang-Mills-Chern-Simons model in 5 dimensions. Such a construction provides a new connection between the  Skyrme model and holographic QCD. In the spirit of this construction we propose a general framework, from which it is possible to obtain both Skyrme and BPS Skyrme models form a supersymmetric Born-Infeld (BI) type action in 5 dimensions.

This work is organized as follows: In Sec. II we propose a general scheme to build SUSY extensions of general bosonic models in $3+1$ dimensions. In Sec. III we apply this scheme to construct several SUSY extensions of the Skyrme model. In Sec. III.B, the proposal of \cite{Sutcliffe} is embedded in a supersymmetric model in 5 dimensions and the low energy bound is obtained from one of the central charges of the superalgebra. Sections IV and V are devoted to studying the SUSY extensions of two Skyrme-like models, the Ferreira- Zakrzewski (FZ) model and BPS Skyrme model. In Sec. VI Skyrme-like models are obtained from a BI-type action and the relation between the Skyrme model and BPS Skyrme model is explained in the context of SUSY BI actions in 5 dimensions. In Sec. VII we present the summary.



\section{General supersymmetric extensions}

We will show in this section that the supersymmetric extension of a given bosonic model is not unique, in the sense that, given a bosonic action it is possible to add the fermionic part in several ways. In order to supersymmetrize the models let us write the following superfield action

\be
\mathcal{L}^S_4=\int d^4\theta \Lambda(\Phi^i, \bar{\Phi}^{i},\Phi,\bar{\Phi} )D_\alpha\Phi D^\alpha\Phi\bar{D}_{\dot{\alpha}}\bar{\Phi}\bar{D}^{\dot{\alpha}}\bar{\Phi}
\label{l4}
\ee
where

\bea
\Phi^i&=&z^i+i\theta \sigma^\mu \bar{\theta}\pa_\mu z^i+\frac{1}{4}\theta\theta\bar{\theta}\bar{\theta}\square z^i+\sqrt{2}\theta \psi^i-\frac{i}{\sqrt{2}}\theta\theta\pa_\mu\psi^i\sigma^\mu\bar{\theta}+\theta\theta F^i\\
\Phi&=&z+i\theta \sigma^\mu \bar{\theta}\pa_\mu z+\frac{1}{4}\theta\theta\bar{\theta}\bar{\theta}\square z+\sqrt{2}\theta \psi-\frac{i}{\sqrt{2}}\theta\theta\pa_\mu\psi\sigma^\mu\bar{\theta}+\theta\theta F
\eea
$\Phi^i$ and $\Phi$ are chiral superfields and $\Lambda$ is a general function of these superfields and its space-time derivatives. These kinds of supersymmetric terms were used in \cite{Ovrut1} in the context of supersymmetric Galileons and in \cite{Ovrut2} applied to ghost condensates. 

As was pointed out in \cite{Nitta}, it is possible to construct a $N=1$ SUSY extension of any single field model. We will see that this statement can be extended to a set of chiral superfields. The quartic term in superderivatives in (\ref{l4}) saturates the integration over the Grassmann coordinates, this means that the bosonic sector  of the model (\ref{l4}) has the following form

\be
\Lambda(z^i, \bar{z}^i,z,\bar{z})\int d^4\theta D_\alpha\Phi D^\alpha\Phi\bar{D}_{\dot{\alpha}}\bar{\Phi}\bar{D}^{\dot{\alpha}}\bar{\Phi}\vert_{\psi=0}
\ee
i.e., the bosonic sector of the model is determined by the lowest $\theta$ component of $\Lambda$ and the highest $\theta$ component of the quartic term in superderivatives. Its components can be written as follows,

\be
 \mathcal{L}^S_4\vert=\Lambda(z^i, \bar{z}^i,z,\bar{z} )\left(  \vert \pa_\mu z\pa^\mu z\vert^2+2F\bar{F}\pa_\mu \bar{z}\pa^\mu z+(F\bar{F})^2   \right)
 \label{quar}
\ee
where $\vert$ means that we are taking all fermions to zero. The function $\Lambda$ depends on the set of superfields $\Phi^i$ and $\Phi$, and will be chosen in such a way that the bosonic part of the Lagrangian generates the corresponding model. We can add a K\"ahler potential depending on the superfield $\Phi$,

\be
\mathcal{L}^S_2\vert=\int d^4 \theta K(\Phi,\Phi^{\dagger})\vert=g(z,\bar{z})\left( \pa_\mu z\pa^\mu \bar{z}  +F\bar{F}\right)\label{quad}
\ee
where the K\"ahler metric $g(z,\bar{z})$ is given by

\be
g(z,\bar{z})=\frac{\pa^2 K}{\pa z \pa \bar{z}}.
\ee

 We could consider also prepotential terms, but they are useless to our purposes. We are going to consider two families of models, those consisting only of the Lagrangian $\mathcal{L}^S_4$ and those consisting of $\mathcal{L}^S_2+\mathcal{L}^S_4$.


\subsection{Family $\mathcal{L}^S_4$}

Our goal is to prove that it is possible to construct $N=1$ supersymmetric extensions for any bosonic theory consisting of $n$ fields and in particular, that we can obtain BPS equations for the supersymmetric variations of the fermions. Let us consider a general bosonic Lagrangian consisting of a set on $n$ complex fields $\{z^1,...,z^n, \bar{z}^1,...,\bar{z}^n \}$, let us call it $\mathcal{L}(z^1,...,z^n, \bar{z}^1,...,\bar{z}^n )$, and consider the following supersymmetric Lagrangian

\be
\mathcal{L}^S_4(\Lambda)\vert= \Lambda(z^i, \bar{z}^i,z,\bar{z})\left(  \vert \pa_\mu z\pa^\mu z\vert^2+2F\bar{F}\pa_\mu \bar{z}\pa^\mu z+(F\bar{F})^2   \right)
\ee
(which corresponds to the superfield Lagrangian (\ref{l4})). By using the equations of motion (EOMs) for the auxiliary field $F$ we obtain two solutions

\bea
F\bar{F}&=&0\label{can1}\\
F\bar{F}&=&-\pa_\mu z\pa^\mu \bar{z}.\label{can2}
\eea

If we take the first solution $F=0$, the resulting Lagrangian is

\be
\mathcal{L}^S_4(\Lambda)\vert= \Lambda(z^i, \bar{z}^i,z,\bar{z}) \vert \pa_\mu z\pa^\mu z\vert^2
\ee

Now for 

\be
\Lambda(z^i, \bar{z}^i,z,\bar{z} )=\frac{\mathcal{L}(z^1,...,z^n, \bar{z}^1,...,\bar{z}^n )}{\vert \pa_\mu z\pa^\mu z\vert^2}
\ee
we obtain trivially that the bosonic sector of $\mathcal{L}_4$ with this choice of $\Lambda$ constitutes automatically a supersymmetric extension for the theory $\mathcal{L}(z^1,...,z^n)$ in the branch $F=0$, i.e.

\be
\mathcal{L}^S_4=\int d^2\theta d^2\bar{\theta} \frac{\mathcal{L}(\Phi^1,...,\Phi^n, \bar{\Phi}^1,...,\bar{\Phi}^n)}{( \pa_\mu \Phi\pa^\mu \Phi)( \pa_\mu \bar{\Phi}\pa^\mu \bar{\Phi})}
D_\alpha\Phi D^\alpha\Phi\bar{D}_{\dot{\alpha}}\bar{\Phi}\bar{D}^{\dot{\alpha}}\bar{\Phi},\label{bps4}
\ee
where the supersymmetric Lagrangian $\mathcal{L}(\Phi^1,...,\Phi^n, \bar{\Phi}^1,...,\bar{\Phi}^n)$ is the Lagrangian of the bosonic theory where all fields are promoted to chiral/antichiral superfields, keeping space-time derivatives intact. The supersymmetric extension of the theory which corresponds to the other branch (\ref{can2}) has the following form:

\be
\mathcal{L}^S_4=\int d^2\theta d^2\bar{\theta} \frac{\mathcal{L}(\Phi^1,...,\Phi^n, \bar{\Phi}^1,...,\bar{\Phi}^n)}{( \pa_\mu \Phi\pa^\mu \Phi)( \pa_\mu \bar{\Phi}\pa^\mu \bar{\Phi})-( \pa_\mu \Phi\pa^\mu \bar{\Phi})^2}
D_\alpha\Phi D^\alpha\Phi\bar{D}_{\dot{\alpha}}\bar{\Phi}\bar{D}^{\dot{\alpha}}\bar{\Phi}.
\ee

Note that the superfield $\Phi$ is completely absent in the bosonic sector of the theory after the substitution of one solution of the auxiliary field. From here we can try to obtain some information about the BPS equation directly from supersymmetry transformations. The static supersymmetric variations for the fermions are given by

\be
\begin{bmatrix} \delta \psi_\alpha \\ \delta\bar{\psi}_{\dot{\beta}}\end{bmatrix}=\mathcal{M}\cdot \begin{bmatrix} \xi_\alpha\\ \bar{\xi}_{\dot{\beta}}\end{bmatrix}
\ee

\noindent
where

\be
\mathcal{M}=\left( \begin{matrix}  -i\pa_3 z & -i(\pa_1-i\pa_2)z &F &0\\  -i(\pa_1+i\pa_2)z & \pa_3 z&0 &F\\ \bar{F} & 0 & -i\pa_3 \bar{z} & -i(\pa_1-i\pa_2)\bar{z} \\ 0&\bar{F} &  -i(\pa_1+i\pa_2)\bar{z} & \pa_3 \bar{z}
   \end{matrix}\right).
\ee

BPS equations preserving  a part of supersymmetry are generated by the condition $\delta \psi=0$. Then we need to look for some combinations of fermionic transformations giving such a vanishing condition. This is equivalent to the condition $det \mathcal{M}=0$ which gives  the following equation for the auxiliary field

\be
\bar{F}F=-\pa_i z \pa_i \bar{z}\pm\sqrt{ \vert \pa_i z\pa_i z\vert^2-( \pa_i z\pa_i \bar{z})^2  }\label{ferm}.
\ee

Equation (\ref{ferm}) can be seen as an equation for the auxiliary field. After substituting the two solutions for the auxiliary field we arrive at two first order equations depending only on the field $z$,

\bea
\vert \pa_i z \pa_i z\vert&=&0,\quad F\bar{F}=0\label{bps1}\\
\vert \pa_i z \pa_i z\vert-\pa_iz \pa_i \bar{z}&=&0,\quad F\bar{F}=-\pa_iz \pa_i \bar{z}
\eea

Despite the fact that we succeed with the general supersymmetric extension, the variation of fermions only generates trivial BPS equation for the field $z$, which in our construction plays the role of an auxiliary field, since it is absent in the bosonic action. In the following section we consider more possibilities to get information about the BPS equations from supersymmetry.


\subsection{Family $\mathcal{L}^S_2+\mathcal{L}^S_4$}

 By adding (\ref{quar})+(\ref{quad}) and solving the EOMs for the auxiliary field we get

\bea
F\bar{F}&=&0\label{aux0}\\
F\bar{F}&=&-\frac{g(z,\bar{z})}{2 \Lambda(z^i, \bar{z}^i,z,\bar{z}) }-\pa_\mu z\pa^\mu \bar{z}\label{aux11}.
\eea

If we consider the branch (\ref{aux0}) we will arrive again at (\ref{bps4}) plus the term corresponding to the K\"ahler potential. Let us take then the solution (\ref{aux11}). For the total Lagrangian ($\mathcal{L}^S=\mathcal{L}_2^S+\mathcal{L}_4^S$) we get

\be
\mathcal{L}^S(\Lambda)\vert=\Lambda(z^i, \bar{z}^i,z,\bar{z} )\left( \vert \pa_\mu z\pa^\mu z\vert^2-( \pa_\mu z\pa^\mu \bar{z})^2  \right)-\frac{g(z,\bar{z})^2}{4\Lambda(z^i, \bar{z}^i,z,\bar{z} ) }.
\ee

Now, as we did before, we consider a general bosonic Lagrangian $\mathcal{L}(z^1,...,z^n, \bar{z}^1,..., \bar{z}^n)$, and choose the function $\Lambda$ such that 

\be
\mathcal{L}^S(\Lambda)\vert=\mathcal{L}(z^1,...,z^n, \bar{z}^1,..., \bar{z}^n)
\ee
or equivalently

\be
\Lambda(z^i, \bar{z}^i,\pa_\mu z^i, \pa_\mu\bar{z}^i )=\frac{\mathcal{L}(z^1,...,z^n, \bar{z}^1,..., \bar{z}^n)\pm\sqrt{g(z,\bar{z})^2q-\mathcal{L}(z^1,...,z^n, \bar{z}^1,..., \bar{z}^n)^2}}{2q}
\ee
where $q=\vert\pa_\mu z \pa^\mu z  \vert^2-(\pa_\mu z\pa^\mu\bar{z})^2$. Finally if we promote all complex fields to superfields we obtain the supersymmetric Lagrangian

\bea
\mathcal{L}^S&=&\int d^2\theta d^2\bar{\theta} K(\Phi,\bar{\Phi})+\int d^2\theta d^2\bar{\theta} \frac{\mathcal{L}(\Phi^1,..,\Phi^n,\bar{\Phi}^1,..,\bar{\Phi}^n)}{2Q}D^\alpha \Phi D_\alpha\Phi \bar{D}^{\dot{\alpha}}\bar{\Phi}\bar{D}_{\dot{\alpha}}\bar{\Phi}\label{general}\nonumber\\
&\pm&\int d^2\theta d^2\bar{\theta} \frac{\sqrt{(\pa^2_{\Phi\bar{\Phi}}K(\Phi,\bar{\Phi}))^2Q-\mathcal{L}(\Phi^1,..,\Phi^n,\bar{\Phi}^1,..,\bar{\Phi}^n)^2}}{2Q}D^\alpha \Phi D_\alpha\Phi \bar{D}^{\dot{\alpha}}\bar{\Phi}\bar{D}_{\dot{\alpha}}\bar{\Phi}\nonumber\\
&&
\eea
where $Q =(\pa_\mu\Phi\pa^\mu\Phi)(\pa_\mu\bar{\Phi}\pa^\mu\bar{\Phi})-(\pa_\mu \Phi\pa^\mu\bar{\Phi})^2$. This Lagrangian verifies, as we stated before the following property:

\be
\mathcal{L}^S\vert_{\psi=0,\psi^i=0,on-shell}=\mathcal{L}(z^1,...,z^1, \bar{z}^1,...,\bar{z}^n)
\ee
i.e., $\mathcal{L}^S$ is the on-shell $N=1$ supersymmetric extension of any theory consisting of $n$ complex fields, once we use the solution (\ref{aux11}) for the auxiliary field. Now we can analyze what we obtain from the supersymmetric variations of the fermions. By substituting the static part of (\ref{aux11}) in (\ref{ferm}) we arrive at

\be
g(z,\bar{z})\sqrt{q^2}\mathcal{L}(z^1,...,z^1, \bar{z}^1,...,\bar{z}^n)\vert_{static}=0\label{genBPS}
\ee
which is a first order equation provided that $\mathcal{L}(z^1,...,z^1, \bar{z}^1,...,\bar{z}^n)\vert_{static}$ is of first order. We will see later applications of this result. We have shown therefore that there exist at least two nonequivalent SUSY extensions for any bosonic theory consisting of $n$ complex fields.

In some cases, the analysis of the BPS structure of the model can be done in a simpler way. Let us take the following family

\be
\mathcal{L}=\int d^4 \theta \left(K(\Phi^i,\bar{\Phi}^i)+\sum_{i=1}^n\Lambda_i (\Phi^j)D^\alpha\Phi^i D_\alpha \Phi^i \bar{D}^{\dot{\alpha}}\bar{\Phi}^i \bar{D}_{\dot{\alpha}}\bar{\Phi}^i  \right) 
\ee

We impose the following conditions on the K\"ahler potential

\be
\frac{\pa^2K}{\pa \phi^i \pa\bar{\phi}^j }=0, \quad for \quad i\neq j
\ee

These models are a restriction over the ones studied in \cite{Nitta,Nitta2}. For other models with similar structure see \cite{Farakos1,Farakos2}. The function $\Lambda$ is a general function depending only on the $n$ superfields $\Phi^1,...,\Phi^n$. Note that we drop the dependence of the extra superfield $\Phi$, and the K\"ahler potential depends on the fields $\Phi^i$.  After integration over the Grassmann space we get

\be
\mathcal{L}\vert_{\psi=0}=\sum_{i=1}^n\left \{ g_{i,\bar{i}}\left( \pa_\mu z^i \pa^\mu \bar{z}^{\bar{i}}+F^i\bar{F}^{\bar{i}} \right)+\Lambda_i(z^j)\left(\vert \pa_\mu z^i \pa^\mu z^i \vert^2  +2 F^i\bar{F}^{\bar{i}} \pa_\mu z^i \pa^\mu \bar{z}^{\bar{i}} +( F^i\bar{F}^{\bar{i}})^2\right)\right \}
\ee

After solving the equations of motion for $F^i\bar{F}^i$ we obtain again two possibilities

\bea
F^i\bar{F}^i&=&0\label{canF}\\
F^i\bar{F}^i&=&-\pa_\mu z^i\pa^\mu\bar{z}^{\bar{i}}-\frac{g_{i,\bar{i}}}{2\Lambda_i (z^j)}\label{noncanF}.
\eea

From the canonical branch (\ref{canF}) we obtain the following bosonic sector

 \be
\mathcal{L}\vert_{\psi=0}=\sum_{i=1}^n\left \{ g_{i,\bar{i}} \pa_\mu z^i \pa^\mu \bar{z}^{\bar{i}}+\Lambda_i(z^j)\vert \pa_\mu z^i \pa^\mu z^i \vert^2  \right \}
\ee
and from the supersymmetric variations of fermions we get the following set of trivial first order equations

\be
\vert \pa_\mu z^i \pa^\mu z^i \vert=0,\quad i=1,..,n
\ee

From the second solution (\ref{noncanF}) we obtain

\be
\mathcal{L}\vert_{\psi=0}=\sum_{i=1}^N\left \{ \Lambda_i (z^j)\left( (\pa_\mu z^i\pa^\mu z^i)(\pa_\mu \bar{z}^{\bar{i}}\pa^\mu \bar{z}^{\bar{i}})-(\pa_\mu z^i\pa^\mu \bar{z}^{\bar{i}})^2   \right) -\frac{g_{i,\bar{i}}^2}{4\Lambda_i (z^j)}     \right \}
\ee

Now, in a lump configuration we assume that $\pa_3 z^j=0$. The energy density can be written as

\be
\mathcal{E}=\sum_{i=1}^N\left \{  \left(\sqrt{\Lambda_i(z^j)}\epsilon^{lm}\pa_l\bar{z}^{\bar{i}}\pa_m z^i\pm\frac{g_{i,\bar{i}}}{2\sqrt{\Lambda_i(z^j)}}   \right)^2\mp  	g_{i,\bar{i}}\epsilon^{lm}\pa_l\bar{z}^{\bar{i}}\pa_m z^i\right \}
\ee
where $l,m=1,2$, and therefore

\be
\mathcal{E}\geq \vert g_{i,\bar{i}}\epsilon^{lm}\pa_l\bar{z}^{\bar{i}}\pa_m z^i \vert.
\ee

The bound is saturated when

\be
\epsilon^{lm}\pa_l\bar{z}^{\bar{i}}\pa_m z^i=\pm\frac{g_{i,\bar{i}}}{2\Lambda_i(z^j)}  
\ee
which corresponds to the Bogomolny equations of the model. The first interesting observation is that the bound depends only on the K\"ahler metric, and it is therefore completely independent of the function $\Lambda_i(z^j)$. The second is that, from the supersymmetric variation of fermions we obtain the set of Bogomolny equations. It is easy to see this after substituting (\ref{noncanF}) in (\ref{ferm}) and taking $\pa_3 z^j=0$. Note that if the model has only one complex field and $\Lambda=1/(1+\Phi\bar{\Phi})^4$ the model above represents the $N=2$ supersymmetric extension of the baby Skyrme model \cite{queiruga} when we dimensionally reduce to 3 dimensions.

In general we can extend all these results to lower dimensions.
If we use dimensional reduction from $3+1$ dimensions to $2+1$ dimensions, the supersymmetry of models with $N=1$ is enlarged to $N=2$ in the reduced dimension. Therefore we can conclude that all models consisting of complex superfields have at least one $N=2$ completion. \\
For gauged models the strategy is the same but including vector superfield dependence in the function $\Lambda$. For example, for

\be
\Lambda(\Phi,V)=\frac{D^2\bar{D^2}\left(W^\alpha D_\alpha V+\bar{W}^{\dot{\alpha}}D_{\dot\alpha}V  \right)}{( \pa_\mu \Phi\pa^\mu \Phi)( \pa_\mu \bar{\Phi}\pa^\mu \bar{\Phi})},
\ee 
the Lagrangian

\be
\mathcal{L}=\int d^2\theta d^2\bar{\theta} \Lambda(\Phi,V) D^\alpha \Phi D_\alpha\Phi \bar{D}^{\dot{\alpha}}\bar{\Phi}\bar{D}_{\dot{\alpha}}\bar{\Phi}
\ee

\noindent
generates the $U(1)$ Yang-Mills theory for the branch $F=0$, i.e.

\be
\mathcal{L}\vert_{F=0}=\frac{1}{2}D^2-\frac{1}{4}F^{\mu\nu}F_{\mu\nu}
\ee

We can sum up all the results above in the following corollaries\\ \\
\noindent
\textbf{Corollary 1}: All models in $3+1$ dimensions consisting of $n$ complex fields have at least  one  on-shell supersymmetric completion to $N=1$.\\
\textbf{Corollary 2}: All models in $2+1$ dimensions consisting of $n$ complex fields have at least  one  on-shell supersymmetric completion to $N=2$.


\section{SUSY Skyrme model from instantons}

The action of the Skyrme model can be written as follows:

\be
S=\int d^4x \left( \frac{f_\pi^2}{4}tr (U^{-1}\pa_\mu U)^2+\frac{1}{32e^2}tr [U^{-1}\pa_\mu U,U^{-1}\pa_\nu U]^2 \right)
\ee

\noindent
the parameter $f_\pi$ is identified with the pion decay constant, $e$ is a dimensionless parameter and $U$ is an $SU(2)$ valued field. We can construct a SUSY extension of this model based on the strategy we presented before. Another possibility to obtain an ``approximate" SUSY extension of the Skyrme model consists of supersymmetrizing a 5D Yang-Mills theory, and as was shown by \cite{Sakai} and \cite{Sutcliffe}, the Skyrme model arises as a truncation of this theory. We will see, therefore, that it is possible to supersymmetrize in a natural way a model which contains the Skyrme model an a infinite number of vector fields that we neglect in the last step, despite the fact that after such truncation we break the supersymmetry (in this sense we use``approximate" SUSY extension). We will see that in this case, the Skyrme field is defined as the holonomy of the Yang-Mills field along $z$-lines (where $z$ is the extra dimension) ,\cite{Atiyah},\cite{Sutcliffe}.


\subsection{SUSY Skyrme model: Case 1}

First of all we need to write the action in terms of chiral superfields. Let us define the following matrix of superfields

\be
\mathcal{U}=
\left(
\begin{matrix}
\Phi^1 & -\bar{\Phi}^2 \\
\Phi^2 & \bar{\Phi}^1
\end{matrix}\right)
\ee
plus the constraint $\Phi^i\bar{\Phi}^i=1$. With this definition we observe that when we take $\theta=0$, $\mathcal{U}$ is exactly a $SU(2)$ field. Let us define the following functions in terms of superfields:

\bea
L(\mathcal{U})&=&\frac{f_\pi^2}{4}tr (\mathcal{U}^{-1}\pa_\mu \mathcal{U})^2+\frac{1}{32e^2}tr [\mathcal{U}^{-1}\pa_\mu \mathcal{U},\mathcal{U}^{-1}\pa_\nu \mathcal{U}]^2\\
Q(\Phi)&=&(\pa_\mu\Phi\pa^\mu\Phi)(\pa_\nu\bar{\Phi}\pa^\nu\bar{\Phi})-(\pa_\mu \Phi\pa^\mu\bar{\Phi})^2\\
P(\Phi)&=&(\pa_\mu\Phi\pa^\mu\Phi)(\pa_\nu\bar{\Phi}\pa^\nu\bar{\Phi)}\\
\Sigma_4(\Phi)&=& D^\alpha \Phi D_\alpha\Phi \bar{D}^{\dot{\alpha}}\bar{\Phi}\bar{D}_{\dot{\alpha}}\bar{\Phi}
\eea

The following Lagrangians constitute a $N=1$ extension of the Skyrme model for the $\mathcal{L}^S_4$ family:

\be
\mathcal{L}_{Skyrme}^1=\int d^4\theta L(\mathcal{U})P^{-1}(\Phi)\Sigma_4(\Phi)
\ee
in the branch $F=0$ and

\be
\mathcal{L}_{Skyrme}^2=\int d^4\theta L(\mathcal{U})Q^{-1}(\Phi)\Sigma_4(\Phi)
\ee
in the branch $F\bar{F}=-\pa_\mu z\pa^\mu \bar{z}$. And for the family $\mathcal{L}^S_2+\mathcal{L}^S_4$ we have

\be
\mathcal{L}_{Skyrme}^3=\int d^4\theta\left\{ K(\Phi,\bar{\Phi}) +(L(\mathcal{U})-\pa^2_{\Phi\bar{\Phi}}K(\Phi,\bar{\Phi})\pa_\mu\Phi\pa^\mu\bar{\Phi})P^{-1}(\Phi)\Sigma_4(\Phi)\right \}
\ee
in the branch $F=0$ and

\be
\mathcal{L}_{Skyrme}^4=\int d^4\theta\left\{ K(\Phi,\bar{\Phi})+\frac{1}{2}\left(L(\mathcal{U})\pm (\pa^2_{\Phi\bar{\Phi}}K(\Phi,\bar{\Phi})Q(\Phi)-L(\mathcal{U}^2))^{1/2}\right)Q^{-1}(\Phi)\Sigma_4(\Phi)\right \}
\ee
in the branch $F\bar{F}=-\pa_\mu z\pa^\mu \bar{z}$. It is interesting to note that the form of these SUSY actions (especially $\mathcal{L}_{Skyrme}^3$) is closed to the nonlinear action for the chiral Goldstone multiplet \cite{Tseytlin1}-\cite{Park}. 

The Lagrangians $\mathcal{L}_{Skyrme}^a,a=1,2,3,4$ are nonequivalent genuine $N=1$ supersymmetric extensions of the Skyrme model. In such supersymmetric extension the chiral superfield $\Phi$ plays the role of an auxiliary field, since it is completely absent in the bosonic sector for the corresponding branch, but reappears in the fermionic sector. The fact that we have an extra field in the full theory can be avoided by changing the dependence of the functions $P,Q$ and $\Sigma_4$ with one of the fields $\Phi^i$ of the model. But the full theory (with fermions) will contain terms with higher order derivatives in time and thus losing the Hamiltonian interpretation. In order to avoid this problem, in the following section we look for another theory containing the Skyrme model in some limit but whose SUSY completion is well behaved.


\subsection{SUSY Skyrme model as a truncation of SUSY 5D Yang-Mills: Case 2}

Let us start with the five-dimensional maximally supersymmetry $SU(2)$ Yang-Mills theory. The action is given by

\bea
S=&&-\frac{1}{g^2_{YM}}\int d^5 z tr\left( \frac{1}{4}F_{\mu\nu}F^{\mu\nu}+\frac{1}{2}D_\mu X^I D^\mu X^I-\frac{i}{2} \bar{\Psi}\Gamma^\mu D_\mu\Psi \right.\label{instanton}\\
&&\left. \frac{1}{2}\bar{\Psi}\Gamma^5 \Gamma^I \lbrack X^I,\Psi \rbrack-\frac{1}{4}\sum_{I,J}\lbrack X^I,X^J \rbrack^2\right).\nonumber
\eea
where $\mu=0,1,2,3,z$ and $I=1,2,3,4,5$.
It can be shown that the commutator of two supercharges can be written in the following way \cite{Lambert}:

\be
\lbrace   Q_\alpha,Q_\beta\rbrace=P_\mu(\Gamma^\mu C^{-1})_{\alpha\beta}^{-}+Z_5 (\Gamma^5 C^{-1})_{\alpha\beta}^{-}+...
\ee

The dots represent $X^I$-dependent terms that we can neglect (since the Skyrme model will be generated by the pure Yang-Mills part). The term $Z_5$ corresponds to one of the central charges (the only which survives after setting $X^I=0$). It has the following form

\be
Z_5=-\frac{1}{8 g^2_{YM}}\int d^3 x dz tr \left(  F_{ij}F_{kl}\epsilon_{ijkl} \right)\label{central5}.
\ee

We will see now the relation of this supersymmetric model in 5 dimensions with the Skyrme model. There are two approaches to see how the Skyrme model arises from (\ref{instanton}), \cite{Sakai}, \cite{Sutcliffe}. For simplicity we will follow the second one. The gauge fields $A_I$ are $su(2)$-valued, while the bosonic fields are in the fundamental representation (we set these fields to zero). It can be shown that, in a gauge where $A_z=0$, the rest of gauge fields can be written as follows \cite{Sutcliffe}

\be
A_i=-\pa_i U U^{-1}\psi_+ (z)+\sum_{n=0}^{\infty}W_i^n(x)\psi_{n}(z)
\ee
where $\psi_n(z)$ are a basis of Hermite functions defined by

\be
\psi_n(z)=\frac{(-1)^n}{\sqrt{n! 2^n\sqrt{\pi}}}\exp(\frac{z^2}{2})\frac{d^n}{dz^n}\exp[-z^2]\label{hermite}
\ee
and $\psi_+$ 

\be
\psi_+(z)=\frac{1}{2}+\frac{1}{\sqrt{\pi}}\int_0^{z/2}e^{-\xi^2}d\xi
\ee

 $U$ is a $SU(2)$ field defined by the holonomy of $A_z$ along the $z$ direction

\be
U(x)=\mathcal{P} exp \int_{-\infty}^{\infty}A_z (x,z)dz
\ee
and the quantities $W^n_i$ constitute an infinite tower of vector fields. By substituting these expressions we obtain the following field strength 

\bea
F_{zi}&=&-\pa_i U U^{-1}\pa_z \psi_+(z)+\sum_{n=0}^\infty W^n_i(x)\pa_z\psi_n(z)\label{f1}\\
F_{ij}&=& [\pa_i U U^{-1},\pa_j U U^{-1}] \psi_+(\psi_+-1)+\sum_{n=0}^\infty F^n_{ij}(W(x)\psi_n(z))\label{f2}
\eea
where

\be
F^n_{ij}(W(x)\psi_n(z))=\sum_{n=0}^\infty \left( (\pa_i W_j^n(x)-\pa_j W_i^n(x))\psi_n(z)+[W_i^n(x),W^j_n(x)]\psi_n^2(z)\right).
\ee

In we substitute expressions (\ref{f1}) and (\ref{f2}) in (\ref{instanton}) we get the full supersymmetric model. In order to see the emergence of the Skyrme model from this supersymmetric model we switch off $X^I$ fields, vector fields $W^n_i(x)$ and fermions in (\ref{instanton}),

\be
W^n_i(x)=0,\quad X^I=0,\quad \Psi=0.
\ee

The energy we obtain from this model corresponds to the energy of the Skyrme model

\be
E_S=\frac{1}{g_{YM}^2}\int \left( -\frac{c_1}{2}Tr (R_i R_i)-\frac{c_2}{16} Tr ([R_i,R_j]^2) \right) d^3x \label{Skyr}
\ee
where $c_1,c_2$ are certain positive constants and $R_i=\pa_i U U^{-1}$. The instanton number of this Euclidean Yang-Mills theory corresponds to the energy bound of the Skyrme model which is identified with the baryon number. The SUSY extension of the model contains exactly this quantity as the $Z_5$ central charge of the superalgebra (\ref{central5}), 

\bea
Z_5=-\frac{1}{g_{YM}^2}\int d^3x dz&&\left( \epsilon^{ijk}\pa_i U U^{-1}\pa_j U U^{-1} \pa_k U U^{-1}\pa_z\psi_+(z)\psi_+(z) (\psi_+(z)-1)\right.\nonumber\\ 	
&&\left.+f(W^n_i(x))\right)
\eea

If we neglect the vector fields $W^n_i(z)=0$ and integrate over the $z$-direction we obtain finally

\be
Z_5=\frac{1}{6g_{YM}^2}\int d^3x \epsilon^{ijk}\pa_i U U^{-1}\pa_j U U^{-1} \pa_k U U^{-1}
\ee
which corresponds to the baryon number of the Skyrme model.


\section{SUSY quartic Skyrme-like model and Bogomolny bounds: FZ model and SUSY YM}

 Let us take now the following Skyrme-like model proposed by Ferreira and Zakrzewski  \cite{ferreira}:

\be
S=\int d^4 x (\frac{m^2}{2}A^2_\mu-\frac{1}{4e^2}F^2_{\mu\nu})
\ee
where:

\be
A_\mu=\frac{i}{2}(\bar{z}_a\pa_\mu z_a-z_a\pa_\mu\bar{z}_a),\quad F_{\mu\nu}=\pa_\mu A_\nu-\pa_\nu A_\mu\quad  \text{and}\quad a=1,2\label{gaugeFZ}.
\ee

The original fields of this model live on the sphere $\mathbb{S}^3$ ($\bar{z}_a z_a=1$). The Bogomolny equation for static solutions is given by:

\be
\frac{1}{2}\epsilon_{ijk}F_{jk}=\pm meA_i
\ee

The static energy has a lower bound:

\be
E\ge 4\pi^2\frac{m}{3}\vert Q\vert
\ee
with

\be
Q=\frac{1}{4\pi^2}\int d^3 x\epsilon^{ijk} A_iF_{jk}\label{charge}
\ee


\subsection{SUSY formulation of the model: Case 1}

For simplicity we will work now with the family $\mathcal{L}^S_4$, defined in Sec. II.A (but it is also possible to build a supersymmetric extension of the model based on the family $\mathcal{L}^S_2+\mathcal{L}^S_4$) as we did for the case of Skyrme model. First of all we need to define again the superfield analogues of connection and curvature,

\bea
\mathcal{A}_\mu&=&\frac{i}{2}\left( \bar{\Phi}^a \pa_\mu \Phi^a -\Phi^a\pa_\mu \bar{\Phi}^a \right),\quad a=1,2\\
\mathcal{F}_{\mu\nu}&=&i\left( \pa_\mu\bar{\Phi}^a \pa_\nu \Phi^a-\pa_\nu\bar{\Phi}^a\pa_\mu\Phi^a  \right),\quad a=1,2
\eea 

\noindent
for chiral superfields $\Phi^a$, obeying the constraint $\Phi^a \bar{\Phi}^a=1$. Following the strategy we presented before, the action below constitutes an $N=1$ extension of the FZ model in the canonical branch ($F=0$),

\be
S_{FZ}=\int d^4\theta\left( m^2 \mathcal{A}_\mu\mathcal{A}^\mu-\frac{1}{4e^2}\mathcal{F}_{\mu\nu}\mathcal{F}^{\mu\nu}\right)P^{-1}(\Phi)\Sigma_4(\Phi)
\ee
but again, it is interesting to look for another supersymmetric extension free of pathologies in the fermionic sector. In the next section we construct a natural extension and we determine the topological charge from supersymmetry.


\subsection{SUSY formulation of the model: Case 2}
The $N=1$ SUSY version of this model is almost straightforward; the quartic term comes from the $N=1$ super- Yang-Mills (SYM)

\be
\mathcal{L}_{YM}=\frac{1}{e^2}\int d^2\theta W^\alpha W_\alpha +\frac{1}{e^2}\int d^2\bar{\theta}\bar{W}^{\dot{\alpha}}\bar{W}_{\dot{\alpha}}
\ee
or in components

\be
\mathcal{L}_{YM}=-\frac{1}{4e^2}F_{\mu\nu}F^{\mu\nu}-\frac{i}{e^2}\lambda\sigma^\mu\pa_\mu\bar{\lambda}+\frac{1}{2e^2}D^2,
\ee
while the quadratic has the following form in terms of the vector superfield:

\be
\mathcal{L}_2=\frac{m^2}{2}\int d^4\theta V^2=\frac{m^2}{2}A_\mu A^\mu.
\ee

Now by setting $D=0$ we obtain finally

\be
\mathcal{L}_{YM}+\mathcal{L}_2\vert=\frac{m^2}{2}A_\mu A^\mu-\frac{1}{4e^2}F_{\mu\nu}F^{\mu\nu}
\ee

\noindent
and therefore we have an explicit  $N=1$ SUSY extension. Now we can try to obtain an explicit $N=2$ SUSY extension. The first obstruction is the number of physical fields. In order to obtain an $N=2$ extension we can first begin with a dimensional reduction from $6$ dimensions; our action in this case will be

\be
L=\frac{m^2}{2}A_\alpha A^\alpha-\frac{1}{4e^2}F_{\alpha\beta}F^{\alpha\beta}\quad \text{with}\quad \alpha,\beta=0,1,...,5.
\ee

Now assuming that $\pa_4 A_\alpha=\pa_5 A_\alpha=0$, and taking $A_4:=Re(\phi)$ and $A_5:=Im(\phi)$, we can rewrite

\be
L=\frac{m^2}{2}A_\mu A^\mu-\frac{1}{4e^2}F_{\mu\nu}F^{\mu\nu}+\frac{m^2}{2}\phi^\dagger\phi+\frac{1}{2e^2}\pa_\mu\phi^\dagger\pa^\mu\phi
\ee
where now $\mu,\nu=0,1,..,3$. Therefore, we need at least an $N=1$ vector superfield and a complex chiral superfield. The dimensional reduction suggests that the term corresponding to the complex field must be quadratic, leading us to the standard $N=2$ SYM. Since the two supersymmetry generators of the $N=2$ algebra appear on the same footing, the same must be the case with the fermions $\psi^\alpha$ and $\lambda^\alpha$ belonging to the chiral superfield and vector superfield respectively. This implies (as we see in the dimensional reduction) that the extra field $\Phi$ must be rescaled as $\Phi\rightarrow \Phi/e$. The full Lagrangian with the extended SUSY can be written as

\be
\mathcal{L}_{N=2}=\frac{1}{e^2}(\int d^2\theta W^\alpha W_\alpha +\int d^2\bar{\theta}\bar{W}^{\dot{\alpha}}\bar{W}_{\dot{\alpha}})+\frac{1}{e^2}\int d^2\theta d^2\bar{\theta}\Phi^\dagger e^{-2V}\Phi
\ee
or in components

\be
\mathcal{L}_{N=2}=-\frac{1}{4e^2}F_{\mu\nu}F^{\mu\nu}+\frac{1}{e^2}(D_\mu\phi)^\dagger D_\mu\phi+\frac{1}{e^2}(\frac{1}{2}D^2-D\phi^\dagger\phi+F^\dagger F)+(fermions) \label{l2}.
\ee

It is also possible to add a Fayet-Iliopoulos term

\be
\mathcal{L}_{FI}=\xi\int d^4\theta V.
\ee



Our auxiliary fields are now $D$ (from the vector superfield) and $F$ (corresponding to the extra field $\Phi$). From the EOMs we obtain

\bea
F&=&0\\
D&=&\phi^\dagger\phi-e^2\xi
\eea

Rewriting the on-shell Lagrangian:

\be
\mathcal{L}=-\frac{1}{4e^2}F_{\mu\nu}F^{\mu\nu}+\frac{1}{e^2}(D_\mu\phi)^\dagger D_\mu\phi-\frac{1}{2}\left( \frac{1}{e}\phi^\dagger \phi-e\xi  \right)^2+(fermions)
\ee

Now since the field $\phi$ comes from the dimensional reduction and it is not relevant for our purposes, we can eliminate it by taking the vacuum constant solution, 

\be
\phi=\exp[i\eta] e \sqrt{\xi}
\ee
(Note that if the Fayet-Iliopoulos term is absent, $\phi$ has a vanishing vacuum value.) In order to evaluate the central charge we can determine the anticommutator between supercharges from the full supersymmetric Lagrangian (see \cite{gaume}),

\be
\lbrace Q_{(1),\alpha},Q_{(2),\beta}\rbrace=-\frac{2\sqrt{2}}{e^2}\epsilon_{\alpha\beta}\int d^3 x(i F^{0i}+\tilde{F}^{0i})D_i\phi^\dagger
\ee 
but for static configurations $F^{0i}=0$, while $\tilde{F}^{0i}=\frac{1}{2}\epsilon^{ijk}F_{jk}$. If we take the constant vacuum solutions for $\phi$ we obtain finally the central charge of the algebra

\be
\lbrace Q_{(1),\alpha},Q_{(2),\beta}\rbrace=\frac{i\sqrt{2}}{e^2}\epsilon_{\alpha\beta}\int d^3 x\epsilon^{ijk}F_{jk}A_i
\ee
which is nothing but the topological charge given in expression (\ref{charge}).


\section{SUSY BPS Skyrme model in 3+1 dimensions}

The Lagrangian density for the BPS Skyrme model in 4 dimensions can be written as follows:

\be
\mathcal{L}_{BPS}=-\lambda^2\pi^2B_\mu B^\mu-\mu^2 V(U,U^\dagger)
\ee
where the dynamical variable $U$ takes values on $SU(2)$. The topological current can be expressed as

\be
 B^\mu=\frac{1}{24\pi^2}\epsilon^{\mu\nu\rho\sigma}tr (R_\nu R_\rho R_\sigma)
 \ee
 where $L_\mu=U^\dagger \pa_\mu U \in su(2)$. The usual parametrization for $U$ \cite{adam1}, consists of a real scalar field $\xi$ and 3-component unit vector $\hat{n}$:
 
 \be
 U=e^{i\xi \hat{n}\cdot \tau}
 \ee

The condition $\det U=1$ is automatically satisfied  since $tr ( \hat{n}\cdot \tau)=0$ ($\tau$ are the Pauli matrices). After the stereographic projection

\be
\hat{n}=\frac{1}{1+\vert u\vert^2}\left( u+\bar{u},+i(u-\bar{u}),\vert u\vert^2-1  \right) 
\ee
the Lagrangian density for the model can be written in terms of a complex field $u$ and a real scalar field $\xi$, 

\be
\mathcal{L}=\frac{\lambda^2 sin^4\xi}{(1+\vert u\vert^2)^4}\left(  \epsilon^{\mu\nu\rho\sigma}\xi_\nu u_\rho\bar{u}_\sigma   \right)^2-\mu^2 V(\xi,u,\bar{u})\label{lagreal}
\ee

In terms of these variables the Bogomolny equation can be written as follows

\be
\frac{\lambda \sin^2 \xi}{\left( 1+\vert u \vert^2  \right)^2}\epsilon^{jkl} i \xi_j u_k \bar{u}_l =\mp \mu\sqrt{V}
\ee
where $a_i\equiv \pa_i a$. Despite the fact that this parametrization is useful in a lot of situations, it does not provide the most natural field content in a supersymmetric extension of the model. We can start with two complex fields $(z^1,z^2)$ and one constraint $(z^1,z^2)\in\mathbb{S}^3$. The field $U$ can be written as

 \[ U=\left( \begin{array}{cc}
z^1 & -\bar{z}^2   \\
z^2 & \bar{z}^1  \end{array} \right)\] 
provided that $\det U=\vert z^1\vert^2+\vert z^2\vert^2=1$. In terms of the complex variables $z^1,z^2$ the topological current can be written as

\be
B^\mu=\frac{1}{4\pi^2}\epsilon^{\mu\nu\rho\sigma}(\vert z^1\vert^2+\vert z^2\vert^2 )\left( z^1_\nu \bar{z}^1_\rho(\bar{z}^2 z^2_\sigma-z^2\bar{z}^2_\sigma)+z^2_\nu \bar{z}^2_\rho(\bar{z}^1 z^1_\sigma-z^1\bar{z}^1_\sigma)\right)\label{current}
\ee

In the SUSY version of the model the complex fields will be promoted to a pair of chiral/antichiral fields, which are the correct field content of the hypermultiplet in 4 dimensions. Moreover if we define the connection

\be
A_\mu=\frac{i}{2} (\bar{z}^a z^a_\mu-z^a\bar{z}^a_\mu)\label{gauge}
\ee
and the curvature

\be
F_{\mu\nu}=\pa_\mu A_\nu -\pa_\nu A_\mu=i(\bar{z}_\mu^a z_\nu^a-\bar{z}_\nu^a z_\mu^a)
\ee

then the topological current can be expressed in a simpler way,

\be
B^\mu=\frac{1}{4\pi^2}\epsilon^{\mu\nu\rho\sigma}(\vert z^1\vert^2+\vert z^2\vert^2 )F_{\nu\rho}A_\sigma=\frac{1}{4\pi^2}\epsilon^{\mu\nu\rho\sigma}F_{\nu\rho}A_\sigma
\ee
and therefore the Lagrangian density can be written as

\be
\mathcal{L}=-\frac{\lambda^2}{4\pi^2}\left( \epsilon^{\mu\nu\rho\sigma}F_{\nu\rho}A_\sigma\right)^2-\mu^2 V(z^i,\bar{z}^i).
\label{lag}
\ee

From (\ref{lag}) we can obtain again the BPS bound

\be
E\ge\frac{\lambda \mu}{2\pi}\int d^3 x \sqrt{V(z^i,\bar{z}^i)}\epsilon^{ijk}F_{ij}A_k
\ee
and the BPS equation 

\be
\frac{\lambda}{2\pi}\epsilon^{ijk}F_{ij}A_k=\mp \mu \sqrt{V(z^i,\bar{z}^i)}\label{eqBPS}
\ee


\subsection{SUSY formulation of the model: Case 1}

In our first supersymmetric version of the BPS Skyrme model, we follow the procedure presented in the previous sections. First of all we define the following quantities

\bea
\mathcal{A}_\mu&=&\frac{i}{2}\left( \bar{\Phi}^a \pa_\mu \Phi^a -\Phi^a\pa_\mu \bar{\Phi}^a \right)\\
\mathcal{F}_{\mu\nu}&=&i\left( \pa_\mu\bar{\Phi}^a \pa_\nu \Phi^a-\pa_\nu\bar{\Phi}^a\pa_\mu\Phi^a   \right)\\
L&=&-\frac{\lambda^2}{4\pi^2}\left( \epsilon^{\mu\nu\rho\sigma}\mathcal{F}_{\nu\rho}\mathcal{A}_\sigma \right)^2-\mu^2 V(\Phi^a,\bar{\Phi}^a)
\eea

Now taking into account the supersymmetric invariant constraint $\Phi^a\bar{\Phi}^a=1$ and the nontrivial solution for F (\ref{aux11}), the following Lagrangian constitutes a supersymmetric extension of the BPS Skyrme model (\ref{lag}) in the family $\mathcal{L}^S_2+\mathcal{L}^S_4$:

\be
\mathcal{L}_{BPS}=\int d^4\theta\left\{ K(\Phi,\bar{\Phi})+\frac{1}{2}\left(L\pm (\pa^2_{\Phi\bar{\Phi}}K(\Phi,\bar{\Phi})Q(\Phi)-L\right)^{1/2}Q^{-1}(\Phi)\Sigma_4(\Phi)\right \}
\ee

Note that it is possible to construct three extra nonequivalent SUSY extensions, but as we saw in some examples, from the family $\mathcal{L}^S_2+\mathcal{L}^S_4$ we can obtain BPS equations from the supersymmetric variations of the fermions. Applying to the present case the Eq. (\ref{genBPS}) in the static regime we arrive at

\be
g(z,\bar{z})\sqrt{q}\left(\frac{\lambda^2}{4\pi^2}\left(\epsilon^{ijk}F_{ij}A_k\right)^2-\mu^2 V(z^i,\bar{z}^i)  \right)=0
\ee
for $q=Q\vert$. Now for $g(z,\bar{z})\neq0$ and $q^2\neq 0$ this equation corresponds to the square of ($\ref{eqBPS}$) and therefore BPS equation of the model is deduced from the supersymmetric variation of the fermions.

The construction above is explicitly $N=1$ supersymmetric, but since the model has a topological current $B^\mu$ we can ensure that in fact has $N=2$ supersymmetry \cite{Spector}. This can be seen in the following way. First of all, once we have the $N=1$ supersymmetric form of the theory written in terms of the variables $(z^i, \bar{z}^i)$ plus the constraint, we can solve the constraint and write the model in the variables $(\xi,u,\bar{u})$ (\ref{lagreal}). In this situation we can write the topological current as follows:

\bea
J_\mu^{top}&=&\epsilon_{\mu\nu\rho\sigma}\pa^\nu B^{\rho\sigma}\\
B^{\rho\sigma}&=&\frac{i}{2\pi^2}\frac{\xi-\sin[\xi]\cos[\xi]}{\left(1+\vert u \vert^2\right)^2}u^\rho\bar{u}^\sigma
\eea

The specification of the potential $B^{\rho\sigma}$ is not unique since the topological current is invariant under the transformation

\be
B^{\mu\nu}\rightarrow B^{\mu\nu}+\pa^\mu \lambda^\nu\label{gaugeinv}
\ee
for any $\lambda^\nu$. Now following \cite{Spector}, we can place the topological current in a real linear superfield $L$, which can be written in terms of chiral spinor superfields $\Phi^\alpha, \bar{\Phi}^{\dot{\alpha}}$,

\be
L=D^\alpha \Phi_\alpha+\bar{D}^{\dot{\alpha}}\bar{\Phi}_{\dot{\alpha}}
\ee
such that the conservation of the topological current can be obtained from the condition $D^2L=0$ without using the equations of motion. Now due to the gauge invariance (\ref{gaugeinv}), we can choose a gauge where $\pa_\mu B^{\mu\nu}=0$. Taking into account that we placed the potential for the topological current in the linear superfield $L$, we can consider the supersymmetry transformations of the potential and construct two new quantities

\be
\tilde{S}^\alpha_\mu=[ \bar{Q}_{\dot{\alpha}},B_{\mu\nu}(\sigma^\nu)^{\alpha\dot{\alpha}}],\quad \bar{\tilde{S}}^{\dot{\alpha}}_\mu=[ Q_\alpha,B_{\mu\nu}(\sigma^\nu)^{\alpha\dot{\alpha}}]\label{spin}
\ee
but from $\pa_\mu B^{\mu\nu}=0$ we get

\be
\pa^\mu\tilde{S}^\alpha=0,\quad \pa^\mu  \bar{\tilde{S}}^{\dot{\alpha}}_\mu=0
\ee
i.e., they are spinor conserved currents. The original $N=1$ supercurrents verify \cite{Haag}

\be
 \{ Q_\alpha,S^\alpha_\mu  \}=0,\quad  \{ \bar{Q}_{\dot{\alpha}},\bar{S}^{\dot{\alpha}}_\mu  \}=0,
\ee
but it can be shown \cite{Spector}, that the new spinorial supercurrents (\ref{spin}) must verify

\be
 \{ Q_\alpha,\tilde{S}^\alpha_\mu  \}\propto J_\mu,\quad  \{ \bar{Q}_{\dot{\alpha}},\bar{\tilde{S}}^{\dot{\alpha}}_\mu  \}\propto J _\mu
\ee

Therefore these new supercurrents are neither trivial nor the original $N=1$ supercurrents of the model; this implies that they are new supercurrents corresponding to an extended $N=2$ supersymmetric invariance of the theory.

Let us note that although the method presented here allows us to make a systematic construction of supersymmetric extensions of any theory, it leads to some problems hidden in the fermionic sector. Indeed, the terms proportional to $Q^{-1}$ can generate pathologies in the fermion part of the full action.
 It is still interesting then to look for more natural SUSY completions of this model.  In the next section we construct a SUSY extension of the BPS Skyrme model free of these problems.
  

\subsection{SUSY formulation of the model: Case 2}

 Guided by the expression (\ref{lag}) for the Lagrangian we can assume that the supersymmetric version of the theory can be written in terms of a vector superfield (generating the kinetic term) and chiral superfield (generating the potential term). The superfield content of the theory will be

\be
\Phi^1 (z^1,\psi_{\alpha}^1, F^1)\oplus\Phi^2 (z^2,\psi_{\alpha}^2, F^2)\oplus V (A_\mu, \lambda_\alpha, D)
\ee

 Such an $N=1$ theory must be invariant under the following set of transformations:

\bea
\delta z^I&=&\sqrt{2}\xi^\alpha\psi_\alpha^I\\
\delta\psi_\alpha^I&=& i\sqrt{2}\sigma^\mu_{\alpha\dot{\alpha}}\bar{\xi}^{\dot{\alpha}}D_\mu z^I+\sqrt{2}\epsilon_\alpha F^I\\
\delta F^I&=&i\sqrt{2}\bar{\xi}_{\dot{\alpha}}\bar{\sigma}^{\mu\dot{\alpha}\alpha}D_\mu\psi_\alpha^I-2i\bar{\xi}_{\dot{\alpha}}z^I\bar{\lambda}^{\dot{\alpha}}\\
\delta A_\mu &=& i\left( \bar{\xi}_{\dot{\alpha}}\bar{\sigma}^{\mu\dot{\alpha}\alpha}\lambda_\alpha-\xi^\alpha\sigma_{\mu\alpha\dot{\alpha}}\bar{\lambda}^{\dot{\alpha}}  \right)\\
\delta \lambda_\alpha&=& \frac{1}{2}\xi^\beta \sigma^{\mu\nu}_{\beta\alpha}F_{\mu\nu}+i\xi_\alpha D\\
\delta D&=& \bar{\xi}_{\dot{\alpha}}\bar{\sigma}^{\mu\dot{\alpha}\alpha}\pa_\mu\lambda_\alpha-\xi^\alpha \sigma^\mu_{\alpha\dot{\alpha}}\pa_\mu\bar{\lambda}^{\dot{\alpha}}
\eea
where $D_\mu=\pa_\mu+i A_\mu$. We need to find the supersymmetric invariant constraint corresponding to the constraint over the bosonic components of the chiral fields. This can be done by varying with supersymmetric transformations the original condition \cite{nepo}, \cite{lisa}. The following set of supersymmetric invariant constraints is obtained:

\bea
\bar{z}^iz_i&=&1\\
z^i\bar{\psi}_i&=&0\\
z^i\bar{F}_i&=&0\\
A_\mu&=&\frac{i}{2}(\pa_\mu\bar{z}^iz_i-\bar{z}^i\pa_\mu z_i)-\frac{1}{2}\psi^{I\alpha}\sigma_{\mu\alpha\dot{\alpha}}\bar{\psi}^{\dot{\alpha}I}\label{gauge1}\\
\lambda_\alpha&=&-\frac{1}{\sqrt{2}}i \bar{F}^I \psi^I_\alpha+\frac{1}{\sqrt{2}}\sigma^\mu_{\alpha\dot{\alpha}}D_\mu z^I\bar{\psi}^{I\dot{\alpha}}\\
D&=&\bar{D}^\mu\bar{z}^iD_\mu z_i-\bar{F}^iF_i+\frac{i}{2}\left(\psi^{I\alpha}\sigma^\mu_{\alpha\dot{\alpha}}\bar{D}_\mu\bar{\psi}^{I\dot{\alpha}}-D^\mu\psi^{I\alpha}\sigma_{\mu\alpha\dot{\alpha}}\bar{\psi}^{I\dot{\alpha}}\right)\label{aux}
\eea

We observe that the components of the vector superfield are totally determined by the components of the hypermultiplet ($\Phi_1, \Phi_2$); moreover, in the bosonic restriction of the theory, the form of the gauge field $A_\mu$ coincides exactly with (\ref{gauge}), which means that, in principle we can construct the supersymmetric form of the kinetic term in (\ref{lag}) in terms of only the superfield $V$. In order to construct out action besides the gauge field $A_\mu$ we need also the field strength $F_{\mu\nu}$. The natural supersymmetric objects which contain the field strength are the superfield strengths, $W_\alpha$ and $\bar{W}_{\dot{\alpha}}$ defined by

\be
W_\alpha=-\frac{1}{4}\bar{D}\bar{D} D_\alpha V, \quad \bar{W}_{\dot{\alpha}}=-\frac{1}{4}DD \bar{D}_{\dot{\alpha}} V
\ee

From the form of the bosonic Lagrangian (\ref{lag}) we deduce three possible  terms generating combinations with at most six derivatives:

\bea
\alpha_1&=&W^\alpha W_\alpha D^\beta V D_\beta V \\
\alpha_2&=&\bar{W}_{\dot{\alpha}}\bar{W}^{\dot{\alpha}}\bar{D}_{\dot{\beta}}V\bar{D}^{\dot{\beta}}V\\
\alpha_3&=& W_\alpha D^\alpha V \bar{W}_{\dot{\alpha}} \bar{D}^{\dot{\alpha}}V.
\eea

After integration, we get for the bosonic sector

\bea
\frac{1}{4}\int d^2\theta d^2\bar{\theta}\,\alpha_1\vert&=&\left(-\frac{1}{2}F_{\mu\nu}F^{\mu\nu}+D^2+\frac{i}{4}\epsilon^{\mu\nu\rho\sigma}F_{\mu\nu}F_{\rho\sigma}\right) A_\rho A^\rho\\
\frac{1}{4}\int d^2\theta d^2\bar{\theta}\,\alpha_2\vert&=&\left(-\frac{1}{2}F_{\mu\nu}F^{\mu\nu}+D^2-\frac{i}{4}\epsilon^{\mu\nu\rho\sigma}F_{\mu\nu}F_{\rho\sigma}\right) A_\rho A^\rho
\eea
while for $\alpha_3$

\be
\mathcal{L}_3=\frac{1}{4}\int d^2\theta d^2\bar{\theta}\,\alpha_3\vert=\frac{1}{2}D^2 A_\mu A^\mu+\frac{1}{4}F_{\mu\nu}F^{\mu\nu}A_\rho A^\rho +F_{\mu\nu}A_\rho F^{\mu\rho}A^\nu
\ee

Taking into account that

\be
\frac{1}{2}(\epsilon^{\mu\nu\rho\sigma}F_{\nu\rho}A_\sigma)^2=F_{\mu\nu}F^{\mu\nu}A_\rho A^\rho+2F_{\mu\nu}A_\rho F^{\mu\rho}A^\nu
\ee
we disregard possible contributions of $\alpha_1$ and $\alpha_2$. In order to construct the prepotential we will need the following contributions:

\be
\mathcal{L}_P^1=\int d^2\theta d^2\bar{\theta}W_\alpha W^\alpha \bar{W}^{\dot{\beta}}\bar{W}_{\dot{\beta}}\vert=(D^2-\frac{1}{2}F_{\mu\nu}F^{\mu\nu})^2+(\frac{1}{2}\tilde{F}_{\mu\nu}F^{\mu\nu})^2
\ee

(in our case $\tilde{F}_{\mu\nu}F^{\mu\nu}=0$), and

\bea
\mathcal{L}_P^2&=&\int d^2\theta K(Z^I)W^\alpha W_\alpha+\int d^2\bar{\theta}K(\bar{Z}^I)\bar{W}_{\dot{\alpha}}\bar{W}^{\dot{\alpha}}\vert=\\ \nonumber
&=&(K(z^I)+K(\bar{z}^I))(-\frac{1}{4}F_{\mu\nu}F^{\mu\nu}+\frac{1}{2}D^2)\\
\mathcal{L}_P^3&=&\int d^2\theta d^2\bar{\theta} D_\alpha V D^\alpha V \bar{D}_{\dot{\alpha}}V \bar{D}^{\dot{\alpha}}V=\left( A_\mu A^\mu\right)^2\\
\mathcal{L}_P^4&=&\int d^2\theta d^2\bar{\theta} \left(K(\Phi^i)+K(\bar{\Phi}^i)\right)V^2=\left(K(z^i)+K(\bar{z}^i)   \right)A_\mu A^\mu
\eea

Now we can combine all these contributions (they are $N=1$ supersymmetric independently):

\be
\mathcal{L}_{BPS}=\alpha\mathcal{L}_3+\beta_1\mathcal{L}_P^1+\beta_2\mathcal{L}_P^2+\beta_3\mathcal{L}_P^3+\beta_4\mathcal{L}_P^4 
\ee
where

\be
\alpha=-\frac{\lambda^2}{2\pi^2},\quad \beta_1=1,\quad \beta_2=\sqrt{\mu},\quad\beta_3=\frac{\lambda^4}{64\pi^4}\quad \text{and}\quad \beta_4=-\frac{\lambda^2}{16\pi^2}\sqrt{\mu}
\ee
and after solving the equation of motion of the auxiliary field D, [note that this fix the equation for $F^i\bar{F}^i$ (\ref{aux})],

\be
D^2=\frac{1}{2}F_{\mu\nu}F^{\mu\nu}-\frac{\sqrt{\mu}}{4}(K(z^i)+K(\bar{z}^i))+\frac{\lambda^2}{8\pi^2}A_\mu A^\mu
\ee
we arrive at

\be
\mathcal{L}_{BPS}=-\frac{\lambda^2}{4\pi^2}\left( \epsilon^{\mu\nu\rho\sigma}F_{\nu\rho}A_\sigma\right)^2-\frac{\mu^2}{16} (K(z^i)+K(\bar{z}^i))^2
\ee
which corresponds to the $N=1$ SUSY extension of the BPS Skyrme model (\ref{lag}) for 

\be
V(z^i,\bar{z}^i)=\frac{ (K(z^i)+K(\bar{z}^i))^2}{16}.
\ee

\section{Skyrme models and BI action in 5D}

Non-Abelian BI action can provide a good description of the dynamics of $N$ coincident D-branes, since its linearized version
is nothing but $SU(N)$ Yang-Mills theory. This action can be written in $d$-spatial dimensions as follows:

\be
\mathcal{L}_d=\mathcal{T}_d Tr\left[ I-\sqrt{-\det\left(\eta_{ab}I+\frac{1}{T} F_{ab}\right)}\right]\label{BIgen}
\ee
with $\mathcal{T}_d$ the brane tension and $T$ a parameter measuring the nonlinearity of the theory (which is related to the string tension) . In the $1/T$-expansion  of (\ref{BIgen}), the first term is exactly the Yang-Mills term; therefore, in the context of Sakai-Sugimoto construction \cite{Sakai}, this action contains the Skyrme model. Our goal in the next sections will be the study of the relation between the higher order term in such an expansion with other Skyrme models, especially the BPS sextic Skyrme. We will analyze also the  supersymmetry version of the model.


\subsection{FZ model plus sextic Skyrme  from the BI action of the $U(1)$ $D4$ brane}

 Let us assume the Abelian situation; for static configurations of a  D4-brane ($F_{0a}=0$) the above action can be written as \cite{brecher}

\be
\mathcal{L}_4=\mathcal{T}_4 \left(1-\sqrt{1+\frac{1}{4T^2}F^2+\frac{1}{4T^4}\tilde{F}^2+\frac{1}{16T^4}(F\cdot \tilde{F})^2}\right)
\ee
where $F\equiv F_{ab},\, a,b=1,2,3,4$, and the Hodge dual $\tilde{F}$ is taken with respect to space indices. If we expand the square root and disregard terms of higher order we get

\be
\mathcal{L}_4=-\frac{1}{4}\mathcal{T}_4\left( \frac{1}{T^2} F^2+\frac{1}{8T^4}(F\cdot \tilde{F})^2+\frac{1}{8T^4}(F^2)^2+\mathcal{O}\left(\frac{1}{T^5}\right)\right)\label{DBI}
\ee

Let us define $A_4\equiv A_z$ and $x_4=z$. The gauge fields can be expanded in terms of Hermite functions (\ref{hermite}), \cite{Sutcliffe},\cite{Sakai}, in a gauge where $A_I\rightarrow 0$ as $\vert z\vert\rightarrow\infty$,
\be
A_{z}(x,z)=\sum_{n=0}^\infty \alpha^n_i (x)\psi_n(z),\quad A_i(z)=\sum_{n=0}^\infty \beta^n_i(x) \psi_n (z)
\ee

Now considering a gauge transformation $A_a\rightarrow A_a-\pa_a h$, such that $A_z=0$ the gauge transformation can be written as  \cite{Sutcliffe}

\be
h(x,z)=u \frac{1}{\sqrt{2}\pi^{1/4}}\int_{-\infty}^z \psi_0 (\xi)d\xi+\sum_{n=0}^\infty h^n \psi_n(z)
\ee
where $u=h(x,\infty)=\int_{-\infty}^{\infty}A_z(x,\xi)d\xi$.  We can write the field strength as follows:

\bea
F_{zi}&=& -\frac{1}{\pi^{1/4}}\pa_i u \frac{\psi_0(z)}{\sqrt{2}}+\sum_{n=0}^{\infty}v_i^n\pa_z\psi_n(z)\\
F_{ij}&=&\sum_{n=0}^\infty\left( \pa_i v_j^n-\pa_jv_i^n  \right) \psi_n(z)
\eea
where $v_i^n=\beta_i^n-\pa_i h^n$. The model has an infinite tower of vector fields $v_i^n(x)$ and a scalar field $u$. The emergence of Skyrme-like models from the BI action can be seen by keeping one of these vector fields $v_i^0(x)$ and neglecting $v_i^n(x)=0,n\geq 1$. In such situation the field strength can be rewritten like

\bea
F_{zi}&=& -\frac{1}{\sqrt{2}\pi^{1/4}}\pa_i u \psi_0(z)-\frac{1}{\sqrt{2}}v_i^0 \psi_1(z)\label{fiz}\\
F_{ij}&=&\left( \pa_i v_j^0-\pa_jv_i^0  \right) \psi_0(z)\label{fij}
\eea

By substituting (\ref{fiz}) and (\ref{fij}) in (\ref{DBI}) and integrating over z we get the static Lagrangian

\bea
\mathcal{L}_4&=&-\frac{\mathcal{T}_4}{4}\left[ \frac{1}{T^2} \left( \frac{1}{2\sqrt{\pi}}(\pa_i u)^2+\frac{1}{2}a_i^2+f_{ij}^2   \right) +\frac{3}{8\sqrt{2\pi}T^4}\left(a_i a_if_{jk}f_{jk}+2a_k a_i f_{jk}f_{ij}  \right)\right.\nonumber\\
&&\left. \frac{1}{4\pi\sqrt{2}T^4}\left( \pa_i u\pa_i u f_{jk}f_{jk}+2 \pa_k u\pa_i u f_{jk}f_{ij}  \right)  \right]-\frac{1}{8T^4}\left[ \frac{1}{\sqrt{2\pi}}(f_{ij}f_{ij})^2+\right.\nonumber\\
&&\left.+\frac{3}{16\sqrt{2\pi}}(a_ia_i)^2+\frac{1}{4\sqrt{2}\pi^{3/2}}(\pa_i u \pa_i u)^2+\frac{1}{4\pi\sqrt{2}}a_ia_i\pa_ju\pa_ju    +\right.\nonumber\\
&&\left.+\frac{1}{2\sqrt{2}\pi}a_i\pa_iua_j\pa_ju+\frac{1}{2\sqrt{2\pi}}f_{ij}^2a_k^2+\frac{1}{\pi\sqrt{2}}f_{ij}^2(\pa_ku)^2 \right]   \label{fzbps}
\eea
where $f_{ij}\equiv \pa_i a_j-\pa_j a_i$ and $a_i\equiv v_i^0$. We can identify

\be
a_i=\frac{i}{2}\sum_{a=1}^2\left(\bar{z}^a\pa_i z^a-z^a\pa_i\bar{z}^a \right), \quad \bar{z}^a z^a=1, \quad z^a\in \mathbb{C}
\ee

We can also take the constant solution for the $\sigma$-model field $u$. In this situation the Lagrangian (\ref{fzbps}) takes the form

\bea
\mathcal{L}_4&=&-\frac{\mathcal{T}_4}{4}\left[ \frac{1}{T^2} \left(\frac{1}{2}a_i^2+f_{ij}^2   \right) +\frac{3}{8\sqrt{2\pi}T^4}\left(a_i a_if_{jk}f_{jk}+2a_k a_i f_{jk}f_{ij}  \right)\right]  \label{fzbps1}\\
&&-\frac{1}{8T^4}\left[ \frac{1}{\sqrt{2\pi}}(f_{ij}f_{ij})^2+\frac{3}{16\sqrt{2\pi}}(a_ia_i)^2+\frac{1}{2\sqrt{2\pi}}f_{ij}^2a_k^2 \right] \nonumber
\eea

The Lagrangian (\ref{fzbps1}) corresponds to the sum of the FZ model and the BPS Skyrme model without potential, and one extra four-derivative term.
The full DBI model before the truncation has a low energy bound which can be written in terms of the topological density

\be
\mathcal{E}_4=\mathcal{T}_4 \left(1-\sqrt{(1+\frac{1}{4T}F\cdot{F})^2+\frac{1}{4T}(F-\tilde{F})^2}\right)\geq \frac{\mathcal{T}_4 }{4T} F\cdot \tilde{F}
\ee
or in terms of the expansion

\be
\mathcal{E}_4\geq \frac{\mathcal{T}_4}{4T}\epsilon^{ijk}\left( f_{ij}^0 \left(-\frac{1}{\pi^{1/4}}\pa_k u+v_k^1\right)+\frac{1}{\sqrt{2}}\sum_{n=0}^\infty f_{ij}^n \left(v_k^{n+1}\sqrt{n+1}-v_k^{n-1}\sqrt{n}\right)\right)
\ee
and after the truncation 

\be
\mathcal{E}_4\geq \frac{\mathcal{T}_4}{4\pi^{1/4}T}\vert\epsilon^{ijk}f_{ij}^0\pa_k u\vert
\ee
i.e., the energy bound depends of the scalar field $u$ (for $u=const.$ it is trivial). Therefore the restricted model does not preserve any topological information.








\subsection{Skyrme model plus sextic Skyrme  from the BI action of the $SU(2)$ D4 brane}

The $SU(2)$ BI action for a D4 brane configuration in 5 dimensions can be written as before:

\be
\mathcal{L}_4=\mathcal{T}_4 Tr\left[ I-\sqrt{-\det\left(\eta_{ab}I+ \frac{1}{T} F_{ab}\right)}\right]\label{naDBI}
\ee

where here $F_{ab}$ in an $SU(2)$-valued field strength. In the non-Abelian case we can expand the determinant in the Lagrangian as follows \cite{brecher}:

\be
-\det \left( \eta_{ab} I+ \frac{1}{T}F_{ab} \right)=I+\frac{1}{2 T^2}F^2+\frac{1}{3T^3}F^3+\mathcal{O}\left(\frac{1}{T^4}\right)\label{lagbi}
\ee
where $F^2=F_{ab}F^{ab}$ and $F^3=F_{ab} F^{bc}F_c^a$ (trace over theses expressions is understood) and the indices $a,b,c,d$ take the values $0,1,2,3,z$. Note that in the non-Abelian case there is an ambiguity over the order of the factors in the determinant. This ambiguity can be solved by taking the symmetrized trace \cite{Tseytlin0}. Nevertheless it is possible to include odd powers of $F$ \cite{brecher}, and this is the case at hand. We will see that the inclusion of odd powers of $F$ generates the emergence of the BPS Skyrme model at order $\frac{1}{T^3}$. If we substitute this expression in (\ref{naDBI}) and expand in powers of $F$ disregarding terms of higher order we get

\be
\mathcal{L}_4=-\frac{\mathcal{T}_4}{2} Tr \left( \frac{1}{2T^2}F^2+\frac{1}{3 T^3}F^3+\mathcal{O}(\frac{1}{T^4})\right)\label{totalDBI}
\ee

The first term in the expansion corresponds to an $SU(2)$ Yang-Mills term in 5-dimensional space. We will determine the contribution of the second term in the context of Sakai-Sugimoto construction \cite{Sakai},\cite{Sutcliffe}. As we did in Sec. III.B we rewrite the field strength in terms of the $SU(2)$ field $U$ and the contribution of the vector fields $W^n_i(x)$ [expressions (\ref{f1}) and (\ref{f2})]. From the Yang-Mills term after setting to zero the fields $W^n_i(x)$ the Skyrme model is obtained. The point here is what is generated by the first higher order term. From the terms in the expansion we get (for the static case)

\be
F^2=\left( 2(\pa_z\psi_+)^2 R_i R_i +\psi_+^2(\psi_+-1)^2[R_i,R_j]^2 \right)\label{FF2}
\ee

\bea
F^3&=&\frac{3}{2}Tr[R_i,R_j]^2 \psi_+(\psi_+-1)(\pa_z\psi_+)^2+\nonumber\\
&+&Tr[R_i,R_j][R_j,R_k][R_k,R_i]\psi_+^3(\psi_+-1)^3\label{F3}
\eea
where $R_i =\pa_i U U^{-1}$. The emergence of the Skyrme model from the $F^2$ term and the extra contribution to the quartic term from the $F^3$ term are clear, but we are interested also in the possible emergence of the sextic term of the BPS Skyrme model. This term can be written as

\be
\mathcal{E}_6=\left(Tr \left(\epsilon^{ijk} R_i R_jR_k \right) \right)^2
\ee

 The above expression can be written as a single trace term up to a factor $2$, 

\be
\mathcal{E}_6=2Tr\left( \epsilon^{ijk} R_i R_jR_k   \right)^2=-\frac{3}{2} Tr[R_i,R_j][R_j,R_k][R_k,R_i]
\ee
 which corresponds to the second line of (\ref{F3}). If we substitute (\ref{FF2}) and (\ref{F3}) into (\ref{totalDBI}) and integrate in $z$ we get for the energy functional
 
 \bea
E_{DBI}=-&&\frac{\mathcal{T}_4}{2}\int d^3x \left( \frac{1}{2\sqrt{\pi}T^2}Tr(R_i R_i)+\frac{1}{T^2}\left(\lambda_{21}-\frac{\lambda_{22}}{T}\right) Tr[R_i,R_j]^2+\right.\nonumber\\
&&\left.+\frac{4\lambda_3}{3 T^3} Tr\left(\epsilon^{ijk}R_iR_jR_k)^2\right)+\mathcal{O}\left(\frac{1}{T^4}\right)  \right)
\eea

\begin{table}[h!]
  \begin{center}
    \caption{Numerical values.}
    \label{tab:table1}
    \begin{tabular}{l|c|c}
      $\lambda_{21}$ & $\lambda_{22}$ & $\lambda_{3}$ \\
      \hline
      0.0495 & 0.0276 & 0.0067 \\
    \end{tabular}
  \end{center}
\end{table}

 The numerical values of the constants are shown in Table I. This energy corresponds to the energy of the Skyrme model plus the energy of the BPS Skyrme model without potential. The behavior of the coefficients as functions of the $T$ parameter is shown in Fig. 1. In the limit $T\rightarrow \infty$ we recover the pure Skyrme model, while for the value $T^0=0.558...$ the quartic Skyrme term is canceled and the action has the form of a $\sigma$-model term plus the sextic BPS Skyrme term.
 
 \begin{figure}[h]
    \centering
    \includegraphics[width=0.6\textwidth]{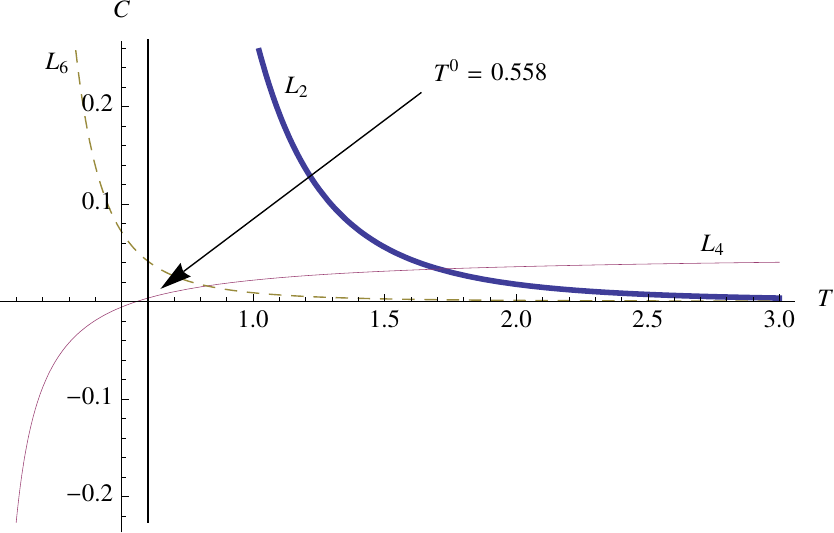}
    \caption{Coefficients $L_2,L_4,L_6$ vs T}
    \label{figmass1}
\end{figure}

If we insist in using the prescription of the symmetrized trace \cite{Tseytlin0}, the term $F^3$ is absent from the Lagrangian in the expansion in $1/T$. In this case the BPS Skyrme term appears at $1/T^4$ order

\bea
\mathcal{L}_4&=&\mathcal{T}_4 STr\left[ I-\sqrt{-\det\left(\eta_{ab}I+ \frac{1}{T}\beta F_{ab}\right)}\right]=\nonumber\\
&=&-\frac{\mathcal{T}_4}{2} Tr\left[  \frac{1}{2T^2}F^2-\frac{1}{16 T^4}\left((F^2)^2-(F\tilde{F})^2\right)+\mathcal{O}\left(\frac{1}{T^6}\right)  \right]
.\label{symnaDBI}
\eea

The term $(F\tilde{F})^2$ will generate the BPS Skyrme term after expansion in modes and integration in z. But in this case another term [$(F^2)^2$], containing higher derivative terms, is generated  [which corresponds to the second line of (\ref{fzbps}) in the Abelian formulation]. We will explain the nature of this extra term in the context of the supersymmetric BI action in the next section.








\subsection{Skyrme and BPS Skyrme from SUSY BI action}










The supersymmetric formulation of the model of the BI model in Euclidean coordinates $\textbf{x},z$ has the following form \cite{Ferrara},\cite{Tseytlin1}:

\be
L_{BI}=\frac{1}{2T^2}\int d^2 \theta \left(W^\alpha W_\alpha+h.c.\right) +\frac{2}{T^4}\int d^4 \theta B(K,\bar{K})W^\alpha W_\alpha\bar{W}^{\dot{\alpha}} \bar{W}_{\dot{\alpha}}\label{SDBI1}
\ee
where

\be
B(K,\bar{K})=\left(  1+\frac{1}{2 T}(K+\bar{K})+\sqrt{1-\frac{1}{T}(K+\bar{K})+\frac{1}{4T^2}(K-\bar{K})^2} \right)^{-1}
\ee
and

\be
K=D^2(W^\alpha W_\alpha),\quad \bar{K}=\bar{D}^2 (\bar{W}^{\dot{\alpha}} \bar{W}_{\dot{\alpha}})
\ee

 The first term in (\ref{SDBI1}) corresponds to the supersymmetric form of the  $SU(2)$ Yang-Mills theory.  If we consider now the expansion of $B$ in powers of $K$



\be
B(K,\bar{K})=\frac{1}{2}+\mathcal{O}(K)
\ee
the action (\ref{SDBI1}) can be written in components as follows

\bea
L_{BI}&=&Tr\left(\frac{1}{T^2}\left( D^a D^a-\frac{1}{2}F^2\right)  +\frac{1}{2T^4}\left(D^aD^a-\frac{1}{2}F^2  \right)^2-\frac{1}{8T^4}(F\tilde{F})^2\right)\\
&+&\mathcal{O}\left( \frac{1}{T^6}  \right) +\text{(fermions)}\nonumber
\eea
where $a$ is the index in $SU(2)$. We can explain now the presence of the term $(F^2)^2$ in the previous models.  If use the trivial solution for $D^a$, $D^a =0$, we obtain the following action

\be
L_{BI}=Tr\left(-\frac{1}{2T^2}F^2 +\frac{1}{8T^4}\left(F^2  \right)^2-\frac{1}{8T^4}(F\tilde{F})^2+\mathcal{O}\left( \frac{1}{T^6}  \right) \right)+\text{(fermions)}\label{BItri}
\ee
which corresponds to the supersymmetric version of the action (\ref{symnaDBI}), and therefore it contains the extra term  $(F^2)^2$. But, we can use the other branch of solutions for the auxiliary fields $D^a$, 

\be
D^a D^a=\frac{1}{2}F^2-T^2
\ee
which gives

\be
L_{BI}=Tr\left(-\frac{1}{8T^4}(F\tilde{F})^2-1/2+\mathcal{O}\left( \frac{1}{T^6}  \right) \right)+\text{(fermions)}
\ee
i.e., the BPS Skyrme term plus a constant. Note that, after the substitution of the auxiliary field, the Yang-Mills term (which generates the Skyrme model term) is completely absent in the action. This situation is quite similar to what happens for the $N=2$ SUSY extension of the baby Skyrme model \cite{queiruga}. We can ask now about the possibility of including a potential term. The natural object to build the potential is the field $U(x)$, which corresponds to the holonomy of $A_z$ in the $z$ direction. The problem is that $U(x)$ does not depend on $z$; therefore, after integration over $z$ in the full model the potential term will a give a vanishing contribution. In order to avoid this problem, we can compactify the $z$ direction in $S^1$, ($\mathbb{M}_5\rightarrow\mathbb{M}_4\times S^1$) and use the Fourier basis in this dimension
 
 \be
 \psi_n(z)=\exp[\frac{2\pi i z}{R}], \quad n=-\infty,..., -1,0,1,...,\infty
\ee
for $z\in [0, R]$ and $R$ the radius of compactification. If $V(U,\bar{U})$ is the potential, its contribution to the energy after $z$-integration will now be $ R V(U,\bar{U})$. The gauge field in the gauge where $A_z=0$ can be written as

\be
A_i=-R_i z+\sum_{n=-\infty}^{\infty}W^n_i(x)\exp[\frac{2\pi i z}{R}]
\ee
and neglecting the contribution of the vector fields $W_i^n(x)$ the field strength can be written as follows:

\bea
F_{zi}&=&-R_i\label{F11}\\
F_{ij}&=&[R_i,R_j]z(z-1)\label{F12}
\eea
 
We introduce now in the SUSY version of the BI action (\ref{SDBI1}) a prefactor $h(\mathcal{U})$ in the Yang-Mills term. This term will play the role of potential as we will see below:

\be
L_{BI}=\frac{1}{2T^2}Im[i\int d^2 \theta \left(h(\mathcal{U})W^\alpha W_\alpha+h.c.\right)] +\frac{2}{T^4}\int d^4 \theta B(K,\bar{K})W^\alpha W_\alpha\bar{W}^{\dot{\alpha}} \bar{W}_{\dot{\alpha}}\label{SDBI2}
\ee

Note that we change the YM term to avoid the $\theta$-term $i (h(U)-h(\bar{U}))F\tilde{F}$. The superfield $\mathcal{U}$ is an $SU(2)$ chiral superfield such that the holonomy $U$ of the gauge field $A_z$ is contained in the lowest component, i.e.  $\mathcal{U}\vert_{\theta=0}=U$. The Lagrangian (\ref{SDBI2}) can be written in components as follows:

\bea
L_{BI}&=&Tr\left(\frac{1}{T^2}(h(U)+h(\bar{U}))\left( D^a D^a-\frac{1}{2}F^2\right)  +\frac{1}{2T^4}\left(D^aD^a-\frac{1}{2}F^2  \right)^2\right.\\
&&-\left.\frac{1}{8T^4}(F\tilde{F})^2\right)
+\mathcal{O}\left( \frac{1}{T^6}  \right) +\text{(fermions)}\nonumber
\eea

We substitute now the solutions for the auxiliary field $D^a$. From $D^a=0$ we obtain (\ref{BItri}) with the prefactor $(h(U)+h(\bar{U}))$ in the YM term (which corresponds to the Skyrme term), the sextic BPS term, and the square of the Skyrme term. Let us analyze now the other solution

\be
D^aD^a=\frac{1}{2}F^2-T^2(h(U)+h(\bar{U}))
\ee

The BI energy in this branch can be written like

\be
\mathcal{E}_{BI}=\int d^3x dz\left( -\frac{1}{8T^4}(F\tilde{F})^2+\frac{1}{2}\left((h(U)+h(\bar{U}))\right)^2  \right)+\text{(fermions)}\label{static5D}.
\ee

If we substitute (\ref{F11}) and (\ref{F12}) in (\ref{static5D}) we finally obtain

\be
\mathcal{E}_{BI}=Tr\int d^3x \left( -\frac{2}{T^4}\gamma(R)\left(\epsilon_{ijk}F_{ij}A_k\right)^2+\frac{R}{2}\left((h(U)+h(\bar{U}))\right)^2  \right)+\text{(fermions)}\label{static5D1}
\ee
where $\gamma(R)=\frac{R^3}{3}-\frac{R^4}{2}+\frac{R^5}{5}$. This is the static energy functional of the BPS Skyrme model with potential $V(U,\bar{U})=\left(h(U)+h(\bar{U})\right)^2$. It is interesting to note that the potential is constructed completely in terms of the holonomy of the gauge field $A_z$. This implies that, in some sense the full SUSY BPS Skyrme model can be constructed in superfield formulation in terms of a single vector superfield in 5 dimensions. The components in the $z$-direction of the field strength correspond to the gauge field $A_i$, the spatial components to the curvature of this gauge field and the holonomy of $A_z$ plays the role of chiral $SU(2)$ superfield.  In our construction this holonomy can be used to build potentials.  The low energy bound can be obtained directly from (\ref{static5D1})

\be
\mathcal{E}_{BI}\geq Tr\frac{2i}{T^2}\int d^3 x\sqrt{R\gamma(R)}\vert\epsilon_{ijk} F_{ij}A_k \vert\sqrt{V(U,\bar{U})}   \label{lowE}
\ee
and the BPS equation

\be
\frac{2i}{T^2}\sqrt{\frac{\gamma(R)}{R}}Tr\epsilon_{ijk}F_{ij} A_k=\pm Tr\sqrt{V(U,\bar{U})}
\ee

In the limit where the radius of compactification tends to zero the BPS furnishes the constant vacuum solutions $Tr V(U,\bar{U})=0$ since $\lim_{R\rightarrow0}\frac{\gamma(R)}{R}=0$ and the low energy bound tends to zero.
On the other hand, in the large radius limit, the energy bound (\ref{lowE}) tends to infinity. The same results are obtained in the $U(1)$ case for the FZ  and BPS Skyrme model from SUSY BI action.

















\section{Summary}

In this work we have studied different aspects of Skyrme-like models and supersymmetry. First, we have obtained a general procedure to build SUSY extensions of any bosonic model consisting of $n$ complex fields in $3+1$ dimensions. Based on this proposal, we have built several  SUSY extensions of the Skyrme, FZ and BPS  Skyrme models. 

We have argued that these general extensions may contain certain pathologies in the fermionic sector, for example the breakdown of the Hamiltonian interpretation. Because of this reason we have built also well-behaved SUSY extensions of all these models. Despite the fact that we were able to construct several SUSY extensions to these models in four dimensions, it seems that, at least for the Skyrme and BPS Skyrme models, 5 dimensions is more natural. In the constructions of \cite{Sakai} and \cite{Sutcliffe}, the emergence of the Skyrme model is closely related to the existence of the extra dimension $z$. Such a feature allows us to place the term $U^{-1}\pa_i U$ in the z-direction of the field strength $F_{zi}$ and construct quadratic and quartic terms from a single Yang-Mills action. In this context, the Abelian formulation of the BPS Skyrme sextic term can be written in the static regime simply like $(F\tilde{F})^2$ (note that this term identically vanishes for $A_\mu$ of the form (\ref{gaugeFZ}) in 4 dimensions). 

We have also obtained the Bogomolny bounds from supersymmetry. For the FZ model, the BPS bound was computed from the anticommutator of supercharges, while for the BPS Skyrme model it was obtained from the supersymmetric variation of the fermions.

Another interesting fact we have obtained in this work is that the SUSY form of a BI type action provides a nice connection between Skyrme and BPS Skyrme models, which appear placed in different branches of the on-shell theory (for different solutions of the auxiliary fields). This result is completely analogous to what happens in $2+1$ dimensions between the $\mathbb{C}P^1$-model and the $N=2$ BPS baby Skyrme model \cite{queiruga}.

Another interesting problem is the determination of the fermionic sectors of these theories, which due to their complexity remain undetermined. With the knowledge of these sectors we could determine the explicit form of the supercharges and calculate the resulting supersymmetric algebra.
This issue is under current investigation. 

To summarize, we have made some important steps towards a better understanding of the supersymmetric classification of the Skyrme-like models in $3+1$ dimensions. We have found explicitly  SUSY extensions and BPS equations from supersymmetry, which for these theories are new results.

{\bf Acknowledgements.}- The author would like to thank Prof. J. S\'anchez-Guill\'en, C. Adam and A. Wereszczynski for useful comments.  This work was supported by  Funda\c{c}\~{a}o de Amparo \`{a} Pesquisa do Estado de S\~{a}o Paulo (FAPESP).



\end{document}